\documentclass{article} 

\usepackage{graphicx}

 \usepackage{epstopdf}
  \usepackage{color}
\usepackage[nottoc,numbib]{tocbibind}
	\usepackage{relsize}
	\usepackage{amsmath}
	\usepackage{amsthm}
	\usepackage{amssymb}
\usepackage[left=1in,right=1in]{geometry}

   \title{A Reconstruction of Quantum Mechanics}
   \author{Simon Kochen}
   
\begin{document}

\newgeometry{top=1cm}
   \maketitle
 \tableofcontents

    \newcommand\cH{\mathcal{H}}
   \newcommand\cL{\mathcal{L}}
   \newcommand\rfl{\rotatebox{90}{$\rfloor$} }
   \newcommand\vp{\phi}
   \newcommand\tr{\mathop{\rm tr}\nolimits}
   \newcommand\Exp{\mathop{\rm Exp}\nolimits}
      \newcommand\Aut{\mathop{\rm Aut}\nolimits} 
	\renewcommand{\thesubsection}{(\roman{subsection})}
   \newcommand\ssn[1]{\subsection{#1}}
   \newcommand\R{\mathbb{R}}
   \newcommand\Ga{\Gamma}
   \newcommand\ga{\gamma}
      \newcommand\Q{\mathbb{Q}}
  \newcommand \lra[1]{\left\langle#1\right\rangle}
   \newcommand\la{\lambda}
   \newcommand\sig{\sigma}
   \newcommand\Om{\Omega}
   \newcommand\om{\omega}
    \newcommand\bpm{\begin{pmatrix}}
    \newcommand\epm{\end{pmatrix}}
   \newcommand\eps{\epsilon}
    \newcommand\bA{\mathbf{A}}
    \newcommand\bB{\mathbf{B}}
    \newcommand\ppsi{\psi} 
  \theoremstyle{definition}
\newtheorem*{defin}{Definition}

\restoregeometry 
\section*{Introduction}
\addcontentsline{toc}{section}{Inroduction}

Almost a century after the mathematical formulation of quantum mechanics, there is still no consensus on the interpretation of the theory. This may be because quantum mechanics is full of predictions which contradict our everyday experiences, but then so is another, older theory, namely special relativity. 
   
   Although the Lorentz transformations initially gave rise to different 
interpretations, when Einstein's 1905 paper appeared it soon led to a 
nearly universal acceptance of Einstein's interpretation.  Why was 
this? Einstein began with the new conceptual principle that time and simultaneity are  relative to the inertial frame, dropping the classical assumption that they are absolute. By then using the linearity of transformations due to the local nature of special relativity and the experimental fact that the speed of light is constant, Einstein was able to derive the Lorentz transformations. Furthermore, by introducing the natural classical notions of state, observable, and symmetry in the new setting, Einstein derived the new dynamical equations to replace the Newtonian equations. This manifestly consistent derivation allowed for a resolution of the apparent paradoxes which confounded the older ether theory, and led to the adoption of Einstein's interpretation by physicists.     
                                                                                           
In this paper, we shall endeavor to use Einstein's approach as a model for deriving and interpreting quantum mechanics. We also start with a new conceptual precept which replaces a classical premise. It is a basic assumption of classical physics that experiments measure pre-existing inherent observables and properties of systems, and any disturbance due to the interaction with the apparatus can be minimized or incorporated into its effect on the observables. By contrast, when we measure a particle's component of spin in a particular direction in a Stern-Gerlach experiment, it is the general belief that we are                                                                                                                                                                                   not measuring a pre-existing property. Rather, it is the interaction of the particle with the magnetic field, which is inhomogeneous in that direction, that creates the value of the spin.  We shall say that such properties are  \emph{relational} or \emph{extrinsic}, as opposed to the intrinsic properties of classical physics.                                                                                                   

That quantum observables and properties take values only upon suitable interactions is, of course, not new to physicists. Bohr, the founder of the Copenhagen interpretation, wrote in \cite{5}: ``The whole situation in atomic physics deprives of all meaning such inherent attributes as the idealization of classical physics would ascribe to such objects.'' This is a radically new consequence of quantum physics that controverts one of the main conceptual assumptions of classical physics, that properties of a physical system are intrinsic.                              

The aim of this paper is to show that a mathematical formulation of this principle allows us to reconstruct the formalism of quantum mechanics. Let us give the basic idea in defining the structure of extrinsic properties, given in Section 1. Every experiment yields a $\sigma$-algebra of measured properties. For instance, in measuring an quantum observable with spectral decomposition $\sum a_i P_i$, the $\sigma$-algebra is generated by the projections $P_i$. It is shown in Section 1 that for quantum experiments the different measured $\sigma$-algebras cannot all be imbedded into a single $\sigma$-algebra. In the case of classical physics, on the other hand, the measured $\sigma$-algebras all sit inside the $\sigma$-algebra $B(\Omega)$ of intrinsic properties of the system, consisting of the $\sigma$-algebra generated by the open sets of the phase space $\Omega$ of the system.
                                                                                                                            
To mathematically treat the extrinsic properties of quantum mechanics, we replace the encompassing $\sigma$-algebra $B(\Omega)$ of properties by a $\sigma$-\emph{complex} $Q$, consisting of the union of all the $\sigma$-algebras of the system elicited by different decoherent interactions, such as measurements.   
                                                                              
This change allows us to define in a uniform and natural manner the concepts of state, observable, symmetry, and dynamics, which reduce to the classical notions when $Q$ is a Boolean $\sigma$-algebra, and to the standard quantum notions when $Q$ is the $\sigma$-complex $Q(\cH)$ of projections of Hilbert space $\cH$. Moreover, we use this approach to derive both the Schr\"{o}dinger equation and the von Neumann-L\"uders Projection Postulate. We also show on the basis of interferometry experiments why $Q$ has the form $Q(\cH)$.

The most noteworthy feature of this reconstruction of quantum mechanics is that the classical definitions of the key physical concepts such as state, observable, symmetry, dynamics, and the combining of systems take on precisely the same form in the quantum case when they are applied to extrinsic properties. 

  In the standard formulation, these concepts take on a strikingly different form from the classical one. In particular, the definition of state as a complex function and the complex form of the Schr\"{o}dinger equation, as opposed to the intuitive, real definitions of classical physics, led Bohr to speak of this formalism as only a symbolic representation of reality.     
                                                                              
One purpose of this approach is to show that once the relational character of properties is accepted, the definitions of the basic concepts of quantum mechanics are as real and intuitive as is the case for classical mechanics. Of course, it is not our intention to dispense with the linear complex Hilbert space in treating problems in physics. The linearity of the Schr\"{o}dinger equation is crucial for solving atomic problems. Our purpose in showing that our intuitive definitions of the notions are equivalent to the standard complex ones is rather to reduce the use of the complex Hilbert space to a technical computational tool, similar to the use of complex methods in classical electromagnetism and fluid mechanics.            
                                                                                    
At first sight the structure of a $\sigma$-complex $Q$ is unusual. Operations   between                                                                                                                                                                                                                                                                                                                                                                            elements of $Q$ are not defined unless they lie in the same Boolean $\sigma$-algebra within~$Q$. That however is the whole point of this structure. Operations are only defined when they make physical sense. This points to the main difference of this approach to that initiated by Birkhoff and von Neumann \cite{3}, and carried forward by Mackey \cite{17}, and Piron \cite{18}, among others. They define the logic of quantum mechanics to be a certain kind of lattice, consisting of the set of projection operators  of Hilbert space. However, Birkhoff and von Neumann \cite{3} already raised the question: 
\begin{quote}
``What experimental meaning can one attach to the meet and join of two given experimental propositions?''      
\end{quote}
That question has never been adequately answered. Varadarajan, in his  book \cite{21} on the lattice approach to quantum mechanics, written some thirty years after the Birkhoff and von Neumann paper, writes: 
\begin{quote}
``The only thing that may be open to serious question in this is [the] assumption \dots which forces any two elements of $\cL$ to have a lattice sum, \dots We can offer no really convincing  phenomenological argument to support this.''     
\end{quote} 
                                             
 Replacing the structure of a complex Hilbert space by an equally mysterious structure of a lattice does not achieve the goal of a transparent foundation for quantum mechanics. What is perhaps surprising is that the far weaker structure of a $\sigma$-complex suffices to reconstruct the formalism of quantum mechanics. Our approach has nevertheless benefited from the lattice approach, especially as delineated in Varadarajan \cite{21}, since theorems using lattices turned out often to have proofs using the weaker $\sigma$-complex structure.                                               
                                                                                                                                                                                                                     
One of the aims of a consistent, logical reconstruction of quantum mechanics is to resolve problematic questions and inconsistencies in the orthodox interpretation, such as the Measurement Problem, the Einstein-Podolsky-Rosen paradox, the Kochen-Specker paradox, the problem of reduction and the von Neumann-L\"uders Projection Rule, and wave-particle duality.  We discuss a resolution of these questions in the context of this reconstruction as they arise in this paper.

 At various points in the paper we consider properties of systems as they are measured by experiments. We are not however espousing an operational view of quantum mechanics. We believe quantum mechanics describes general interactions in the world, independently of a classical macroscopic apparatus and observer. We do not subscribe to the Bohrian view that classical physics is needed to give meaning to quantum phenomena. The interactions we describe using a macroscopic apparatus could apply equally well to appropriate decoherent interactions between two systems in general. (See the discussion in Section 1).  Nevertheless, we refer for the most part to experiments rather than general interactions in order to emphasize that the postulates have operational content and meaning. This has the merit of allowing those who prefer the operational approach to make sense of this reconstruction. 
 
Another point is that since the properties that constitute a $\sigma$-complex correspond to the results of possible measurements, they refer to what in the orthodox interpretation are the properties that may hold as a result of reduction. We do not attempt to discuss the conditions under which reduction or decoherence occurs.  There are discussions in the literature on the conditions under which reduction can occur. For instance, Bohm \cite{4} analyzes the strength of the inhomogeneity of the magnetic field for a successful reduction to occur in the Stern-Gerlach experiment. We consider these as interesting pragmatic questions which lie outside the purview of this paper.      

We have not given a new axiomatization of quantum physics. In fact, there are no axioms in this paper, only definitions of the basic concepts, definitions which are identical with the classical ones. Rather, we have presented a framework that is  common to all physical theories. It is the aim of every theory to predict the probabilities of the outcomes of interactions between systems. Experiments are particular instances of such interactions. An experiment gives rise to a Boolean $\sigma$-algebra of events which reflects an isomorphic $\sigma$-algebra of properties of the system. The different possible experiments yield a family of $\sigma$-algebras, reflecting a family of $\sigma$-algebras properties of the system, whose union we call a $\sigma$-complex. This $\sigma$-complex helps determine the underlying theory, and conversely, a given theory determines the kind of $\sigma$-complex of perperties that arises, but the general structure of a $\sigma$-complex as a union of $\sigma$-algebras is independent of any particular theory.
                                                 
The main aim of the paper is to derive elementary quantum mechanics by applying the natural classical definitions of the physical concepts to extrinsic properties, and then use this derivation to resolve the standard paradoxes and problematic questions. We shall accordingly give only outlines of the proofs of the requisite theorems. To show that we have accomplished the goal of reconstructing the formalism, we shall use the textbook by Arno Bohm \cite{4}. This book has the advantage of explicitly introducing five postulates which suffice to treat the standard topics in quantum theory. We shall specify each of the Bohm postulates as we derive them in the paper.   

To avoid repetition, we shall make the blanket assumption that the Hilbert space $\cH$ that we deal with is a separable complex Hilbert space. The Appendix has a table which summarizes the reconstruction.
                                                                                                                                                                                                                                                                                                                                                        
\section{Properties}                                                                                                        

  Scientific theories predict the probabilities of outcomes of experiments. We recall from probability theory that the individual outcomes of an experiment on a system form the \emph{sample space} $S$. For instance, a Stern-Gerlach experiment which measures the 
  $z$-components of spin for a spin 1 system has the sample space $S = \{s_{-1}, s_0, s_1\}$ corresponding to the three possible spots labeled $s_{-1}, s_0, s_1$ on the screen. An experiment to measure the temperature of water by a thermometer has (an interval of) the real line as sample space.
  
  Out of the elementary outcomes, one forms an algebra of more complex outcomes,  called \emph{events}, consisting of a Boolean algebra $B$ of subsets of the space $S$. The operations of $B$ consist of  union $a \vee b$,  and complementation $a^\bot$ , and all other Boolean operations, such as intersection $a \wedge b$, which are definable from them. If $S$ is finite, then $B$ consists of all subsets of $S$. If $S$ is infinite, then the operation of countable union $\bigvee a_i$ of elements $a_i$  of $B$, is added, and $B$ is called a (Boolean) $\sigma$-algebra. (For the definition of and details about Boolean algebras see Koppellberg \cite{16}.)                  
                           
  The algebra $B$  of events, i.e.\ sets of outcomes, reflects the corresponding structure of properties of the system. For instance, in the above Stern-Gerlach experiment, the sets $\{s_{-1}\}$, $\{s_0\}$, and $\{s_1\}$ correspond to the properties $S_z= -1$, $S_z=0$, and $S_z=1$; the set $\{s_{-1},s_1\}$ corresponds to the property $S_z=-1 \vee S_z=1$ (where $\vee$ denotes `or'), or equivalently, the property $\rfl(S_z=0)$ (where $\rfl$ denotes `not'), and so on. In this case, the Boolean algebra is clearly the eight element algebra. In the case of the above temperature measurement of the water, the elementary outcomes are open intervals of the real line, and the algebra of events is the $\sigma$-algebra of (\emph{Borel}) sets generated by the intervals by complement and countable intersection.                                                                                                                   

Thus, for both classical and quantum physics, every experiment on  a given system $S$ elicits a $\sigma$-algebra of properties of $S$, which are true or false, i.e.\ have a truth value, for the system. 

We come now to a crucial difference between the two theories. In classical physics, we assume that the measured properties of the system already exist prior to the measurement. It may be true that the interaction of the system with the apparatus disturbs the system, but this disturbance can be discounted or minimized. For instance, the thermometer may change the temperature of the water being measured, but this change can be accounted for, and there is no doubt that the water had a temperature prior to the measurement which is approximated by the measured value. The basic assumption is that systems have intrinsic properties, and the experiment measures the values of some them.

The family of intrinsic properties of a system form a Boolean algebra, and in the infinite case a $\sigma$-algebra. For classical physics, one introduces the phase (or state) space, with a canonical  structure. The open sets of $\Omega$ generate a $\sigma$-algebra $B(\Omega)$ of \emph{Borel} sets by complement and countable intersection. The algebra $B(\Omega)$ constitutes the $\sigma$-algebra of intrinsic properties of the system. Since the $\sigma$-algebras of measured properties are aspects of all the intrinsic properties of the system, these different $\sigma$-algebras must all be part of the $\sigma$-algebra $B(\Omega)$. Hence, the union $\cup B$ of all the $\sigma$-algebras arising from possible measurements is embeddable in $B(\Omega)$. In fact, if we assume that every property of the system is, in principle, experimentally measurable then the union $\cup B$ itself forms a $\sigma$-algebra. 

In quantum mechanics, for measurements such as the Stern-Gerlach experiment, physicists do not believe that the value of the spin component $S_z$ exists prior to the measurement. On the contrary, it is the interaction with the magnetic field, inhomogeneous in the $z$-direction, that results in a definite spot, say $s_1$, on the screen, reflecting the value, $S_z=1$ of the spin of the particle.  
  
  This general conviction is, in fact, supported by a theorem, called the Kochen-Specker Paradox. This result showed that the spin component $S_z$ cannot be an intrinsic property of a spin 1 particle. We recall that this result shows that there exist a small number of directions in space (33 suffice) such that any prior assignment of values to the squares of the components of spin in these directions contradicts the condition that $S_x^2+S_y^2+S_z^2=2$, for an orthogonal triple $(x,y,z)$. Since the squares of the components of spin in orthogonal directions commute for a spin 1 system, we may measure them simultaneously for the triple $(x,y,z)$. For instance, the measurement of the observable $S_x^2-S_y^2$, with eigenvalues 1,-1, 0 gives us the value 0 for $S_x^2, S_y^2$, or $S_z^2$, respectively, and 1 for the other two. We shall call such an experiment a \emph{triple experiment on the frame} $(x,y,z)$. 
  
  The operators $S_x^2,S_y^2,S_z^2$  generate an eight element Boolean algebra:
\[
B_{xyz} =   \{0,1,S_x^2,S_y^2,S_z^2,1-S_x^2,1-S_y^2,1-S_z^2\}.\]                                                                                                                                                                                                                                    
The 33 directions give rise to 40 orthogonal triples, and hence 40 Boolean algebras. It is important to note that the Boolean algebras have common sub-algebras. For instance, the algebra $B_{x'y'z}$ of the triple experiment on $(x', y', z)$ has the Boolean algebra $B_z=(0, 1, S_z^2, 1-S_z^2)$ in common with $B_{xyz}$.                      
                                                                                    
The 40 Boolean algebras, and hence their union $\cup B_{xyz}$, cannot be embedded into a single Boolean algebra. We may see this directly from the fact that every Boolean algebra has truth values, i.e.\ a homomorphism onto the Boolean algebra $\{0,1\}$, so that such an embedding would assign values to all the 40 Boolean algebras simultaneously, and hence to the 40 triples $S^2_x, S^2_y, S^2_z$, contradicting the Kochen-Specker theorem.  (For a proof of this theorem, with the 40 triples, see Conway and Kochen  \cite{6}).      
                                                                                                                                 
   The conclusion is that, in general, quantum mechanical properties are not intrinsic to the system, but have truth values created by interactions with other systems. We shall call such interactive or relational properties \emph{extrinsic}. The question now is: what mathematical structure captures the concept of extrinsic properties, to replace the Boolean $\sigma$-algebras that characterize intrinsic properties?     
                                                               
Such a structure must contain all the $\sigma$-algebras that are elicited by experiments. The minimal structure is then clearly the union $\cup B$, where $B$  ranges over all the   $\sigma$-algebras that arise in experiments. Intuitively, we may obtain such a structure by gluing together the $\sigma$-algebras at the ``faces,'' i.e.\ the common sub-$\sigma$-algebras. This structure is the minimal one which contains all the $\sigma$-algebras arising from different experiments. We shall adopt it as embodying the idea of extrinsic properties. We now give the formal definition of this notion.       
                                                                                                                                                                                                                                                 
\begin{defin}\hspace*{-6pt}\footnote{A Boolean $\sigma$-complex is a closely connected generalization of a \emph{partial Boolean algebra} (introduced in Kochen and Specker \cite{13}, and further studied in \cite{14} and \cite{15}).}
 Let $F$ be a family of $\sigma$-algebras. The $\sigma$-complex $Q_F$ based on $F$ is the union $\cup B$ of all $\sigma$-algebras $B$  lying in $F$. \end{defin} 

We shall generally leave the family $F$ implicit, and simply refer to a $\sigma$-complex $Q$. We shall usually deal with $\sigma$-complexes that are closed under the formation of sub-$\sigma$-algebras. We can, in any case, always close a $\sigma$-complex by adding all its sub-$\sigma$-algebras.

The term $\sigma$-complex is based on the notion of a  \emph{simplicial complex} in topology. A simplicial complex is obtained by taking a family of simplices, which is closed under sub-simplices, and gluing together common simplicial faces. $\sigma$-complexes are not just analogous to simplicial complexes, but have a close correspondence, as we now outline. First recall that an \emph{atom} of a Boolean algebra is an element $x$ such that $y \leq x$ (i.e. $x\wedge y = y$) implies $y =0$ or $y = x$. The atoms of a Boolean algebra in a closed Boolean complex define the vertices of a simplex, and the union of these simplices yield a simplicial complex. We may conversely define a Boolean complex from a simplicial complex. The graphs called K-S diagrams in the literature define simplicial complexes of the cooresponding Boolean complexes. Strictly speaking, a simplicial complex is the family of simplices, and their union is called the carrier, so we should really call $F$ the $\sigma$-complex. However, we shall find it convenient and harmless to conflate the two notions of $\sigma$-complex and its carrier.

Let $\cH$ be a Hilbert space. Every set of pair-wise commuting projection operators closed under the operation of orthogonal complement $P^\bot (=1-P)$ and countable intersection $\bigwedge P_i$ forms a $\sigma$-algebra. We form the family of all such $\sigma$-algebras, and name their union, the $\sigma$-complex based on this family, $Q(\cH)$. The $\sigma$-complex  $Q(\cH)$ is the structure in quantum mechanics that replaces the $\sigma$-algebra $B(\Omega)$ of Borel sets of the phase space $\Omega$ in classical mechanics.

We now summarize this discussion of properties in a form that will serve as a template for each of the other concepts we introduce in the 
later sections. We first give the classical form of the concept in terms of the $\sigma$-algebra $B(\Omega)$; then we generalize the 
concept by simply replacing the $\sigma$-algebra by a $\sigma$-complex $Q$; finally, we specialize to quantum mechanics by taking 
$Q$ to be the $\sigma$-complex $Q(\cH)$. It then requires a theorem to show that the resulting concept is equivalent to the standard 
quantum definition on $\cH$. Some of the classical concepts are defined in terms of the phase space $\Omega$, rather than the 
$\sigma$-algebra $B(\Omega)$. We must then give an equivalent definition of the concept in terms of    $B(\Omega\}$. 
\newline 

\noindent\textit{Classical Mechanics}\

 The properties of a system form the $\sigma$-algebra $B(\Omega)$ of Borel sets of the phase space $\Omega$ of the system.
\\ \newline
\noindent\textit{General Theory}\

 The properties of a system form a $\sigma$-complex $Q$.
\\ \newline
\noindent\textit{Quantum Mechanics}\

 The properties of a system form the $\sigma$-complex $Q(\cH)$ of projections of the Hilbert space $\cH$ of the system.       
\\

For a system $S$ with a $\sigma$-complex $Q$, an appropriate interaction with another system, such as a measurement, or, more generally, a decoherent interaction, will elicit a $\sigma$-algebra $B$  in $Q$ of properties that have truth values. We shall call $B$  \textit{the (current) interactive algebra} for the system $S$ in the interaction.      
                                                               
 For instance, $B_{xyz}$ is the interactive algebra in the triple experiment with the frame $(x, y, z)$. Thus, a measurement of the observable $S_x^2 - S_y^2$  has the interactive algebra  $B_{xyz}$. We may also consider an experiment for which the interactive algebra is $B_z = \{0,1,S_z^2,1-S_z^2\}$. For instance, a variant of the Stern-Gerlach experiment with the magnetic field replaced by an inhomogeneous electric field measures the absolute value $|S_z|$ of $S_z$, since the electric field vector is a polar vector. For a spin 1 system this amounts to measuring $S_z^2$. Such an experiment is described in Wrede \cite{22}. 
 
In general, a measurement of the observable with discrete spectral decomposition $\sum a_iP_i$ has as interaction algebra the $\sigma$-algebra generated by the $P_i$'s. The general case, where the observable contains a continuous spectrum, is described in Section 3.

In the triple experiment, the interaction algebra $B_{xyz}$ of the measured system is reflected in the isomorphic eight element algebra of events consisting of the subsets of the three spots on the detection screen.              
                                                                                                                        
This isomorphism is, as we have seen, a general feature of a measurement, but it is also true for any appropriate decoherent interaction. If the state of the combined two interacting systems is  $\sum a_i\vp_i \otimes \psi_i$ at the end of the interaction, then the interaction algebras of the systems are the two $\sigma$-algebras generated by the $P_{\phi_i}$  and the $P_{\psi_i}$, which are isomorphic.  It is important to note that the macroscopic nature of the apparatus plays no role in the classical nature of the interaction algebras as Boolean $\sigma$-algebras. It simply follows from the nature that we attributed to extrinsic properties, that in every appropriate interaction they have the classical structure of a $\sigma$-algebra.
As a consequence, we have no need to (and do not) subscribe to the Copenhagen interpretation, especially espoused by Bohr, that it is necessary to presuppose a classical physical description of the world in order explicate the quantum world.  Quantum properties are not intrinsic, but the appropriate interaction elicits an interaction algebra with the classical structure of a $\sigma$-algebra.

\section{States}

\ssn{\bf Probability Measures} 
The theory of probability (following Kolmogorov) is based on a \emph{probability measure}, a countably additive, [0,1]-valued measure, i.e.\ a function
\[
p:  B \to [0,1]
\]
 with domain $B$  a $\sigma$-algebra, such that $p(1)=1$, and
\[p(\mathsmaller{\bigvee} a_i) = \sum p(a_i) \hbox{ for pair-wise disjoint elements } a_1, a_2,  \dots \hbox{ in }B.\]

  In the case of a measurement on a system $S$, the probability function $p$ gives the probabilities of the $\sigma$-algebra of events, 
  or equally of the measured properties of  $S$.  A physical theory predicts the probabilities of outcomes of any possible experiment, 
  given the present state. This leads to the following concept of a state.                                                                                                                              \\
																																																				
\noindent\textit{Classical Mechanics: $\sigma$-algebra $B(\Omega)$}\

 A \emph{state} of a system with phase space $\Omega$ is a probability measure on the $\sigma$-algebra $B(\Omega)$.
\\ \newline
\noindent\textit{General Theory: $\sigma$-complex $Q$}\
                                                                                             
A \emph{state} of a system with a $\sigma$-complex of properties $Q$ is a map
                                  $p:  Q \to [0,1]$                                                            
such that the restriction $p|B$ of $p$ to any $\sigma$-algebra $B$  in $Q$ is a probability measure on $B$.         
\\ \newline
\noindent\textit{Quantum Mechanics: $\sigma$-complex $Q = Q(\cH)$}  
\     
                                                                  
Assume that $\cH$ has dimension greater than two. There is a one-one correspondence between states $p$ on $Q(\cH)$ and density operators (i.e.\ positive Hermitean operators of trace 1) $w$ on $\cH$ such that 
\[
 p(x) = \tr(wx) \hbox{ for all } x \in Q(\cH).\]

\smallskip
\noindent That a density operator $w$ defines a probability measure $p$ on $Q(\cH)$ is an easy computation. The converse, that a state $p$ defines a unique density operator $w$ on $\cH$, follows from a theorem of Gleason \cite{11}. Gleason's theorem is the affirmative answere to a question of Mackey\cite{17}, in which Mackey asked whether a state on the lattice of projections on $\cH$ defines a unique density operator.  A careful check of Gleason's proof of the theorem shows that, in fact, the stronger theorem stated above is true, and that the lattice operations on non-commuting projections are not needed for the proof.       

 As this result shows, the intuitive and plausible definition of classical states leads, with the change from intrinsic to extrinsic properties, to a similar characterization of quantum states.                                                                                                                          

\ssn{\bf Pure and Mixed States} 

 The set of states on a $\sigma$-complex is closed under the formation of convex linear combinations:  if  $p_1,p_2, \dots$  are states then so is $\sum c_i p_i$,  for positive $c_i$,  with $\sum c_i = 1$. The above one-one correspondence between states of $Q(\cH)$ and density operators is convexity-preserving. The extreme points of the convex set of states of a system are those that cannot be written as a non-trivial convex combination of states of the system.
\\
 
\noindent\textit{Classical Mechanics: $\sigma$-algebra $B(\Omega)$}
                                                                  
A \emph{pure state} of a system is an extreme point of the convex set of all states of the system.
\\

For $B(\Omega)$, a pure state p has the form   $p(s) = \left\{\begin{array}{cl}
																																				1 \mbox{ if } \omega\in s\\
																																				0 \mbox{ if } \omega\notin s\\ 
																																				\end{array}\right. $.
                                                                                                                                                                                           
In other words, the classical pure states correspond to the points in $\Omega$. Thus, the phase space $\Omega$ consists of the pure states, and so is also called the state space. 

Thus, in the classical case all the properties of the system in a pure state are either true or false. As we would expect for intrinsic 
properties, measurements simply find out which measured properties are the case. The general states as mixtures of the pure states can
 then be interpreted as giving the probabilities of the properties which are true. These may be termed epistemic probabilities, based on 
 the knowledge of the actual pure state that subsists.   
\\ 
                                                         
\noindent\textit{General Theory: $\sigma$-complex $Q$} \

A pure state of a system is an extreme point of the convex set of states of the system.
\\ 
 
\noindent\textit{Quantum Mechanics: $\sigma$-complex $Q  = Q(\cH)$}\
        
There is a one-one correspondence between the pure states of a system and rays $[\psi]$ of unit vectors $\psi$ in $\cH$, such that $p(x) = \lra{\psi,x\psi}$.   \\ 
                                                     
For it is easily seen that the pure states correspond to one-dimensional projections $P_\psi$ (with $\psi$ in the image of  $P_\psi$) and $p(x) = \tr(P_\ppsi x) = \lra{\psi,x\psi}$.  As in the classical case, the state space of the system consists of the pure states, and in this case corresponds to the projective Hilbert space of the rays of  $\cH$.

In the quantum case, even the pure states predict probabilities that are not 0 or 1, and so these are not the probabilities of properties that 
already subsist. This is, of course, what we should expect of extrinsic properties. A pure state simply predicts the probabilities of 
properties in possible future interactions, such as measurements. Mixed states are, as in the classical case, mixtures of the pure states. 
However, in this case there is no unique decomposition of a mixed case into pure states. This has led to a traditional difficulty in 
interpreting quantum mixed states. We shall postpone a discussion of our interpretation of mixed states until we have treated conditional 
probabilities in Section 7.
 
\section{Observables}                                                                                                              

Some classical concepts such as observables are defined using the phase space $\Omega$ rather than the $\sigma$-algebra 
$B(\Omega)$. We can, in general, restate these definitions in terms of $B(\Omega)$. The reason for this is that the Stone Duality
 Theorem between Boolean algebras and spaces (and its extension by Loomis to $\sigma$-algebras) assures us that constructions on the 
 algebras have their counterparts on the spaces and vice versa.
                                                                                                                                    
A classical observable is defined as a real-valued function $f: \Omega  \to \R$ on the phase space $\Omega$ of the system. To avoid 
pathological, non-measurable functions, $f$ is assumed to be a Borel function, i.e.\ a function such that $f^{-1}(s) \in B(\Omega)$, for every set $s$
 in the $\sigma$-algebra $B(\R)$ of Borel sets generated by the open intervals of $\R$.
 
The inverse function $f^{-1}: B(\R) \to B(\Omega)$ is easily seen to preserve the Boolean $\sigma$ operations, i.e.\ to be a 
homomorphism. Moreover, as we see below, any such homomorphism allows us to recover the function $f$.

 For our purposes, the advantage of using the inverse function is that it involves only the $\sigma$-algebra $B(\Omega)$ instead of the 
 phase space $\Omega$, allowing us to generalize the definition to a $\sigma$-complex.                  
\\
      
\noindent\textit{Classical Mechanics: $\sigma$-algebra $B(\Omega)$}\
                                                                          
An \textit{observable} of a system with phase space $\Omega$ is a homomorphism
$u: B(\R) \to B(\Omega)$,  i.e.\ a map u satisfying          
\begin{align*}    
                                                  u(s^\bot)&=u(s)^\bot,\\
                                                  u(\mathsmaller{\bigvee} s_i) &= \mathsmaller{\bigvee} u(s_i),    
                                                  \end{align*}
for all $s,s_1,s_2,\dots  in B(\R)$.
\\

There is a one-to-one correspondence between observables u and Borel functions $f: \Omega\to \R$ such that $u = f^{-1}$.

For given the map u we may define the Borel function $f$ by the equation
\[
                            f(x) = \inf\{ y \mid  y\in\Q,\; x\in  u((-\infty ,y])\} .
                            \]
The proof that $f$ has the requisite properties is direct, using the denumerability of the rationals $\Q$ to apply the countable additivity of $u$. (See Varadarajan \cite[Theorem 14]{21}.)                 
\\
              
\noindent\textit{General Theory: $\sigma$-complex $Q$}\
                                                                                       
 An \textit{observable} of a system with $\sigma$-complex $Q$ is a homomorphism
 \[
 u: B(\R)\to  Q.
 \]
Note that the image of $u$ lies in a single $\sigma$-algebra in $Q$.       
\\

\noindent\textit{Quantum Mechanics: $\sigma$-complex $Q = Q(\cH)$}\
                                                                
There is a one-one correspondence between observables  $u: B(\R) \to Q(\cH)$ and                                                                                                      Hermitean operators $A$ on $\cH$, such that, given $u$,  $A=\int \lambda d P_\lambda$, where $P_\lambda= u((-\infty,\lambda])$.\\

Conversely, given a Hermitean operator $A$ on $\cH$, the spectral decomposition  $A=\int \la d P_\la$  defines the observable $u$ as the spectral measure $u(s)= \int_s dP_\la$, for $s\in B(\R)$.  This establishes the one-one correspondence.        
                                                                                                      
It follows easily that if  $u: B(\R) \to Q(\Omega)$ is an observable with corresponding Hermitean operator $A$, then, for the state $p$ with corresponding density operator $w$, the expectation of $u$
\[\Exp_p(u) = \tr(Aw)              .\]
(See Postulates I and II of Bohm \cite{4}.)

The theorem shows the close connection between the measurement of an observable and the interaction algebra of measured properties. For instance, for  the case of a discrete operator $A$, the spectral decomposition $A = \sum a_i P_i$  defines the                        interaction algebra of measured properties  generated by the $P_i$. Conversely, given the interaction algebra of measured properties, its atoms $P_i$ allow us to define, for each sequence of real numbers $a_i $, the Hermitean operator $\sum a_i P_i$ which is thereby measured. In particular, we may in this way associate an observable with values 0 and 1 with every property in $Q(\cH)$. If $A$ is a non-degenerate observable with eigenvalue $\lambda$ belonging to eigenstate $\phi,$ we shall often speak of the property $A=\lambda$ to mean the projection $P_\phi$ which has image the ray of $\phi.$                                                                                                       

\section{Combined Systems}

An essential part of the formalism of physics is the mathematical description of the physical union of two systems. In this section we 
answer the question: what is the $\sigma$-complex of the union $S_1 + S_2$  of two systems with given $\sigma$-complexes $Q_1$ 
and $Q_2$?

 In classical physics, given two systems $S_1$ and $S_2$ with the phase spaces $\Omega_1$ and $\Omega_2$, the phase space of the combined system $S_1 + S_2$ is the direct product space
 $\Omega_1 \times \Omega_2$, whereas for quantum systems with Hilbert spaces $\cH_1$ and $\cH_2$, the Hilbert space of 
$ S_1 + S_2 $ is the tensor product  $\cH_1\otimes \cH_2$. The direct and tensor products are very different   constructions. The 
dimension of the direct product space is the sum of the dimensions of the two factor spaces, whereas the dimension of the tensor product 
is the product of the dimensions of the factor spaces. It is this difference that lies behind the promise of quantum computers.

We have nevertheless to combine these two operations via a single construction on the $\sigma$-complex $Q$. When $Q = B(\Omega)$, we may get a clue to the construction by means of Stone duality for Boolean algebras  and Boolean spaces. The dual of the direct product of two Boolean spaces is the direct sum $B_1 \oplus B_2$ (also called the free product or co-product) of Boolean algebras.  (See Koppelberg \cite[Chapter 4]{16}.)  A similar duality extends to $\sigma$-algebras. (See \cite[ Chapter 5]{16}.) We now use our general principle of defining a concept on a $\sigma$-complex by reducing it to the corresponding concept on its $\sigma$-algebras.
\\

\noindent\textit{Classical Mechanics: $\sigma$-algebra $B(\Omega)$}\

Given two systems $S_1$ and $S_2$ with $\sigma$-algebras $B(\Omega_1)$ and $B(\Omega_2)$, the combined system $S_1 + S_2$ 
has the $\sigma$-algebra $B(\Omega_1) \oplus B(\Omega_2)$. There is a unique space $\Omega_1 \times \Omega_2$ such that $B
(\Omega_1) \oplus B(\Omega_2) \cong B(\Omega_1\times \Omega_2)$.

The isomorphism is a well-known part of Stone Duality. For a proof see Koppelberg \cite[Chapters 4 and 5]{16}.  
\\

\noindent\textit{General Theory: $\sigma$-complex $Q$}\
   
Given two systems $S_1$ and $S_2$ with $\sigma$-complexes $Q_1$ and $Q_2$, the combined system $S_1 + S_2$ has the $\sigma$-complex $Q_1\oplus Q_2$, consisting of the  closure (i.e.\ all the sub-$\sigma$-algebras) of the direct sums $B_1 \oplus B_2$ of all pairs of $\sigma$-algebras $B_1$ and $B_2$ in $Q_1$ and $Q_2$. 
\\

\noindent\textit{Quantum Mechanics: $\sigma$-complex $Q  = Q(\cH)$}\
                                                         
Given the combined system $S_1 + S_2$ with the $\sigma$-complex  $Q(\cH_1)\oplus Q(\cH_2)$, there is a unique Hilbert space $\cH_1\otimes \cH_2$ such that $Q(\cH_1) \oplus Q(\cH_2) \cong Q(\cH_1\otimes\cH_2)$.
\\(See Postulate IVa of Bohm \cite{4}.) \\         
                                                                                       
  We give an outline of the proof when $\cH_1$ and $\cH_2$ have finite dimensions.  It suffices to show that every element of $Q(\cH_1\otimes\cH_2)$ lies in $Q(\cH_1) \oplus Q(\cH_2)$.  The elements of $Q(\cH_1) \oplus Q(\cH_2)$  are generated by the one-dimensional projections $P_{\vp\otimes \psi}$, where $\phi \in\cH_1 $and $\psi\in\cH_2$. We must show that if $\Ga$ is an arbitrary unit vector in $\cH_1\otimes \cH_2$, then $P_\Ga$ lies in $ Q(\cH_1) \oplus Q(\cH_2)$. One definition of the tensor product allows us to think of $\Ga$ as a conjugate-linear map from $\cH_2$ to $\cH_1$. (See Jauch\cite{12}, for example.)  The proof proceeds by induction on the rank of $\Ga$ as such a map. The maps of rank 1 are of the form $P_{\vp\otimes \psi}$, so the basis of the induction is true.
  
  Now suppose $\Ga$ has rank $n$.
  
  The proof is greatly simplified by choosing  suitable orthonormal bases in $\cH_1$ and $\cH_2$ in which to expand $\Ga$. We can construct bases $\{\vp_i\}$ and $\{\psi_i\}$ in $\cH_1$ and $\cH_2$ such that $\Ga = \sum c_i \vp_i\otimes \psi_i$, with the $c_i$ real. (Briefly, $\Ga\Ga^\ast$ and $\Ga^\ast\Ga$ have common strictly positive eigenvalues, say $a_i$, and respective eigenvectors $\vp_i$ and $\psi_i$; it follows that $\Ga= \sum \sqrt{a_i} \vp_i\otimes \psi_i$. See Jauch \cite{12}, for example.)  
       
 Let

\[
\Theta= 
\begin{cases}
	-c_2 \vp_1\otimes \psi_1+c_1\vp_2\otimes \psi_2, &\mbox{ for } n = 2 \\                 		
	c_1  \vp_3\otimes \psi_1+c_2\vp_2 \otimes \psi_3+c_3\vp_1\otimes \psi_3+\sum_{i>3}c_i \vp_i\otimes \psi_1, & \mbox{ for }n > 2 \\
		\end{cases} \]

\hspace{1.45cm}
$
\Delta= \hspace{0.24cm}
	c_1\vp_2\otimes \psi_1 + c_2\vp_1\otimes \psi_2 +\sum_{i \geq 3} c_i\vp_i\otimes \psi_2 
$\\
						
   Then $\Ga,\Theta$, and $\Delta$  are pairwise orthogonal unit vectors. Hence, $P_\Ga, P_\Delta$, and $P_\Theta$ mutually commute, and $P_\Gamma = (P_\Ga\vee P_\Theta) \wedge ( P_\Ga\vee P_\Delta)$.
   
   For $n=2$, let $x_+ = c_2 \Gamma + c_1 \Theta$ and $x_- = c_1\Gamma - c_2\Theta$. For $n>2$, let $x_\pm = \Gamma \pm \Theta$. Also, let $y_\pm = \Gamma \pm \Delta$. Then it is easily checked that the four vectors $x_\pm$  and 
   $y_\pm$   are of rank $n-1$, and $x_+$ and $x_-$ are orthogonal, as are $y_+$ and $y_-$. It follows that $[P_{x_+} , P_{x_-}] = 
   [P_{y_+}, P_{y_-}] = 0$.  Moreover, $P_\Ga\vee P_\Theta = P_{x_+}\vee P_{x_-}$ and $P_\Ga\vee P_\Delta = P_{y_+} \vee
    P_{y_-}$. Hence,  $P_\Ga = (P_{x_+}\vee P_{x_-})\wedge (P_{y_+}\vee P_{y_-})$. Since $P_{x_{+}} , P_{x_-} , P_{y_+}$,
     and $P_{y_-}$  inductively lie in $Q(\cH_1)\oplus Q(\cH_2)$ and each of the pairs $(P_{x_+},P_{x_-}), (P_{y_+},P_{y_-})$, and
      $(P_{x_+}\vee P_{x_-},P_{y_+}\vee P_{y_-})$ lie in a common  $\sigma$-algebra, it follows that $P_\Ga$ lies in $Q(\cH_1)
      \oplus Q(\cH_2)$. The proof provides an algorithm for constructing $x_\pm$  and $y_\pm$.      
    
The uniqueness (up to isomorphism) is a routine consequence of the fact that $Q_1 \oplus Q_2$ is categorically a co-product (see Koppellberg \cite{16} for a proof in the $\sigma$-algebra case).

The infinite dimensional case is discussed in Section 9. 

As an illustration we consider the simplest case of the tensor product $\cH_1 \otimes \cH_2$ of two-dimensional Hilbert spaces, which we may take to represent two spin $\frac{1}{2}$ particles. Each element of $Q(\cH_1)$ (resp. $Q(\cH_2))$ corresponds to the property $s_z\otimes I= \frac{1}{2}$ (resp. $I\otimes s_z=\frac{1}{2})$ for some direction $z.$ For $\Ga$ in $\cH_1 \otimes \cH_2$ we shall identify $P_{x_+}, P_{x_-}, P_{y_+},$ and $P_{y_-}.$

We write the vector $\Ga$ in the diagonal form $c_1\phi_1 \otimes \psi_1+c_2\phi_2 \otimes \psi_2.$ Hence, \[x_{-} = \phi_1 \otimes \psi_1, \ x_{+} = \phi_2 \otimes \psi_2\]
\[y_{+} = (\phi_1 + \phi_2) \otimes (c_1 \psi_1+c_2\psi_2), \ y_{-}= (\phi_1-\phi_2)\otimes (c_1\psi_1-c_2\psi_2)\]
Now $\phi_1$ defines $s_z \otimes I = \frac{1}{2}$ for a direction $z$, and $\psi_1$ defines $1\otimes s_w = -\frac{1}{2}$ in a direction $w$. Thus, $\phi_1 \pm \phi_2$ defines $s_x \otimes I = \pm \frac{1}{2}$ for a direction $x$ orthogonal to $z.$ Also, if we write $c_1=\cos (\mu/2)$, then $c_1\psi_1+c_2\psi_2$ defines $I\otimes s_u=\frac{1}{2}$ in a direction $u$ at an angle $\mu$ from the $w$ direction, and $c_1\psi_1 - c_2\psi_2$ defines $I\otimes s_v = \frac{1}{2}$ in a direcion $v$ at angle $-\mu$ from the $w$ direction. It follows that 
\begin{align*}
	P_\Ga &= (P_{x_{+}} \vee P_{x_{-}}) \wedge (P_{y_{+}} \vee P_{y_{-}})\\
	&= (s_z \otimes I = \frac{1}{2} \leftrightarrow I \otimes s_w = -\frac{1}{2}) \wedge [(s_x\otimes I = \frac{1}{2} \wedge I\otimes s_u = \frac{1}{2}) \vee (s_x \otimes I = -\frac{1}{2} \wedge I \otimes s_v = \frac{1}{2})]
	\end{align*}
In this manner every state in a combined system can be interpreted as a compound proposition about the factors. 

A particularly interesting case is the singleton state $\Ga = \sqrt{\frac{1}{2}}(\phi^+_z \otimes \psi^-_z - \phi^-_z \otimes \psi^+_z),$ (with $s_z\phi^\pm_z = \pm \frac{1}{2} \phi^\pm_z$ and $s_x \psi^\pm_z = \frac{1}{2} \psi^\pm_z$) where 
\begin{align*}
	P &= (P_\Ga \vee P_\Theta) \wedge (P_\Ga \vee P_\Delta)\\
	&= (S_z=0) \wedge (S_x=0)\\
	&= (P_{x_+} \vee P_{x_-}) \wedge (P_{y_+} \vee P_{y_-})\\
	&= (s_z \otimes I = \frac{1}{2} \leftrightarrow I\otimes s_z = -\frac{1}{2}) \wedge (s_x \otimes I = \frac{1}{2} \leftrightarrow I \otimes s_x = -\frac{1}{2}).
\end{align*}
In Section 10(iv) we shall apply this result to the EPR experiment.

This construction of the direct sum generalizes in an obvious way to the direct sum of an arbitrary number of $\sigma$-complexes, representing the union of
    several systems. The above theorems then generalize to:         
     \begin{align*}
     B(\Omega_1)\oplus B(\Omega_2)\oplus\cdots  &\cong B(\Omega_1\times\Omega_2\times \cdots )\\
Q(\cH_1)\oplus Q(\cH_2)\oplus\cdots  &\cong Q(\cH_1\otimes\cH_2\otimes\cdots ).\end{align*}
These general sums are needed in discussing statistical mechanics. It is now routine to define symmetric and anti-symmetric direct sums of $\sigma$-complexes, yielding the corresponding symmetric and anti-symmetric tensor products of Hilbert spaces, needed to deal with identical particles. (See Postulate IVb of Bohm \cite{4}. The spin-statistics connection that Bohm adds can also be added here.)

\section{Symmetries}
    
 As Noether, Weyl, and Wigner showed, observables such as  position, momentum, angular momentum, and energy arise from global 
 symmetries of space and time, and the conservation laws for them arise from the corresponding symmetries of interactions. Other 
 observables arise from local symmetries. In classical physics the symmetries appear as canonical transformations of phase space, and in
  quantum physics they appear as unitary or anti-unitary transformations of Hilbert space. For us they naturally appear as symmetries 
  of a $\sigma$-complex.
  
  \begin{defin}
   An automorphism  of a $\sigma$-complex $Q$ is  a  one-one transformation     $\sigma:  Q\to Q$  of  $Q$ onto $Q$ such that for every $\sigma$-algebra $B$  in $Q$ and all $a, a_1,a_2, \cdots$     in  $B$
 \[
 \sigma(a^\bot)=\sigma(a)^\bot \hbox{ and }\sigma (\mathsmaller{\bigvee} a_i)=\mathsmaller{\bigvee} \sigma(a_i).\]\end{defin}

\noindent\textit{General Theory: $\sigma$-complex $Q$}\
                                                                                       
A \textit{symmetry} of a system with $\sigma$-complex $Q$ is given by an automorphism of $Q$. \\

A symmetry $\sigma$ defines a natural convexity-preserving map $p\to p_\sigma$  on the states of $Q$ by letting $p_\sigma=p \circ \sigma^{-1}$,   i.e.\ $p_\sigma(x) = p(\sigma^{-1}(x))$, for all $x \in Q$.                                                                                                                                   \\

\noindent\textit{Quantum Mechanics: $\sigma$-complex $Q = Q(\cH)$}         \
                                                       
There is a one-one correspondence between symmetries $\sig: Q(\cH) \to Q(\cH)$ and 
unitary or anti-unitary operators $u$ on $\cH$ such that $\sig(x) = uxu^{-1}$, for all $x \in Q(\cH)$. \\

If a state $p$ corresponds to the density operator $w$, then                                          
\[ p_\sig(x) = p(\sig^{-1}(x)) = \tr(wu^{-1}xu) = \tr(uwu^{-1}x),      \]                       
so that the state $p_\sigma$ corresponds to the density operator $uwu^{-1}$.  

It is easy to check that unitary and anti-unitary operators define a symmetry on $Q(\cH)$. For the converse we use a well-known theorem of Wigner. (See Bargmann \cite{1}.) The original theorem of Wigner posits a one-one map of the set of rays of $\cH$ onto itself which preserves the inner product. Uhlhorn \cite{20} was able to weaken this to preserving the orthogonality of rays. As Bargmann states in \cite{1}, the proof he gives of Wigner's theorem may be easily modified to prove Uhlhorn's result. (For a proof see Varadarjan \cite{21}.)

Now assume that $\sig$ is a symmetry of $Q(\cH)$. Then $\sig$ is a one-one map of the set of atoms, i.e.\ one-dimensional projections $P_\psi$, of $Q(\cH)$ onto  atoms of $Q(\cH)$.   In other words, rays $[\psi]$ of $\cH$ are one-to-one mapped onto rays of  $\cH$. Moreover, since $\sigma$-algebras are mapped by $\sigma$  to $\sigma$-algebras, the orthogonality of rays is preserved. The Uhlhorn version of Wigner's theorem then shows there is a unique (up to a multiplicative constant) unitary or anti-unitary map $u$ on  $\cH$ such that $\sigma(x) = uxu^{-1}$.

In the case of classical physics, with $Q = B(\Omega)$, a symmetry is defined by a canonical transformation of the manifold. Every such transformation defines an automorphism of the $\sigma$-algebra $B(\Omega)$. However, the converse is not true. Although the automorphism still defines a continuous map from $\Omega$ to itself, the structure of a $\sigma$-algebra is too weak to recover the canonical structure. It is remarkable that the $\sigma$-complex structure is sufficient to allow one to define the symmetries of the Hilbert space. In that sense, quantum physics allows a more satisfactory reconstruction than classical physics. As Section 8 suggests, we may recover the classical canonical structure  from the quantum structure in the limit of an increasing number of particles.

\section{Dynamics}

Now that we have shown that the symmetries of $Q(\cH)$ are  implemented by symmetries of  $\cH$, we may use time symmetry to introduce a dynamics for systems.

To define dynamical evolution, we consider systems that are invariant under time translation. For such systems, there is no absolute time, only time differences. The change from time 0 to time $t$ is given by a symmetry $\sig_t:  Q\to  Q$, since the structure of the system of properties is indistinguishable at two values of time. We assume that if the state evolves first for a time t and then the resulting state for a time $t'$, then this yields the same result as the original state evolving for a time $t+t'$. Moreover, we assume that evolution over a small time period results in small changes in the probability of properties occurring. 

 The passage of time is thus given by a continuous representation of the additive group $\R$ of real numbers into the group $\Aut(Q)$ of automorphisms of $Q$ under composition:  \newline i.e. a map  $\sig :\R\to \Aut(Q)$,  such that 
 \[
\sig_{t+t' }= \sig_t \ \mathsmaller{\mathsmaller{\circ}} \ \sig_{t'}\]
and $p_{\sig_t}(x) $ is a continuous function of $t$.          

 The image of $\sig$ is then a continuous one-parameter group of automorphisms on $Q$. \footnote{The group $\Aut(Q)$ may, in fact, be construed as a topological group by defining, for each $\eps>0$,  an $\eps$-neighborhood of the identity to be $\{ \sig\mid |p_\sig(x)- p(x)| < \eps$  for all $x$ and $p \}$. We may then directly speak of the continuity of the map $\sig$, in place of the condition that $p_{\sig_t}(x)$ is continuous in $t$.}
 
We have seen that an automorphism $\sig$ corresponds to a unitary or anti-unitary operator. Anti-unitary operators actually occur as symmetries, for instance in time reversal. However, for the above representation only unitary operators $u_t$  corresponding to the symmetry $\sigma_t$ can occur, since $u_t = u_{t/2}^2$, which is unitary \footnote{More precisely, we have a projective unitary representation of  $\R$, but such a representation of $\R$ is equivalent to a vector representation. (See, e.g., Varadarajan \cite{21}.)}.        

 It follows that the evolving state $p_{\sigma_t}$ corresponds to the density operator                $w_t = u_t wu_t^{-1}$.                                                                                                                                        By Stone's Theorem,
 \[
 u_t = e^{-\frac{i}{\hbar}  Ht} ,        
 \]  
 where $\hbar$ is a constant to be determined by experiment; so
 \[
 w_t = e^{-\frac{i}{\hbar} Ht} w \ e^{\frac{i}{\hbar} Ht}.\]
 Differentiating,           
 \[
 \partial_t w_t = -\frac{i}{\hbar} [ H, w_t ].\]
 This is the Liouville-von Neumann Equation.
 
 Conversely, this equation yields a continuous representation of $\R$ into $\Aut(Q(\cH))$.
  For $w = P_\psi $, a pure state, $w_t  = P_{\psi(t)}$ and this equation reduces to the Schr\"{o}dinger Equation: 
 \[
 \partial_t \psi(t) =-\frac{i}{\hbar} H \psi(t).\]                                     
(See Postulate Va of Bohm \cite{4}. Postulate Vb is the Heisenberg form of the equation, and follows similarly.)

We stop here without specifying any further the form of the Hamiltonian $H$. This form depends upon calculating the linear and angular momentum observables as operators from the homogeneity and isotropy of space, using the corresponding unitary representations that we have used for time homogeneity.
 This a well-known part of quantum mechanics and need not be explored further here. (See Jauch \cite{12}, for example.) We have treated the non-relativistic dynamical equation. The connection between automorphisms of $Q(\cH)$ and unitary operators given above allows to us to treat the relativistic dynamical equations in  a similar manner, following Wigner's work.  (See Varadrajan \cite{21}.)                                                

\section{Reduction and Conditional Probability}   

\ssn{\bf Conditional States}

With these results, which cover four of Bohm's five postulates, we can now recover much of quantum theory. So far however, we will never predict interference. The states we introduced are probability measures on $Q$, which for any experiment is a classical probability measure on the $\sigma$-algebra of properties being measured. In fact, the probability must be classical, since it is mirrored in the probability measure on the experiment's $\sigma$-algebra of events, which are generated by macroscopic spots on a screen.

How then does interference enter the picture? In dealing with experiments, we have omitted a key ingredient that is usually referred to as ``the preparation of state.'' To calculate the probability $p(x)$ of a property holding at the end of an experiment, we need to know both the property $x$ and the state $p$. In general, when we are presented with a particle to be measured, we do not know its state.  One way to know the state is to prepare it by means of  a prior  interaction.  

For instance, the book \cite{8} by Feynman, Leighton, Sands introduces quantum mechanics via  a spin 1 system by discussing the  probability 
of, for instance, going to state $S_x=1$, given that it is in state $S_z =0$. The particle is prepared in state $S_z =0$ by sending it
through a Stern-Gerlach field in the z direction, and then filtering it through a one-slit screen to allow only the central beam through.
 If the system is not detected as hitting the filtering screen, then it is reduced to the state $S_z =0$. If allowed to hit a final detection screen it is certain to register the central spot. But we are free to send it through another Stern-Gerlach field in the $x$ direction to
  measure $S_x=1$, say. This is a reduction by preparation of the original, possibly unknown, state to the state $S_z = 0$.
                                                                                                                                                                                                                                                                                                                        
 Some physicists think that reduction is a phenomenon unique to quantum mechanics that has no counterpart in classical mechanics, but this not the case. Consider a one slit experiment with bullets. If we shoot at a target, we get a probability distribution on the target that defines a mixed state for the bullet. Since the target screen can be placed anywhere from the gun to any distant point, the probability distribution is a function of time that gives a time evolution of this state, satisfying the classical Liouville equation for mixed states. If we now interpose a one-slit screen between the gun and the target screen, we find that after the evolution of the state $p$ up to the one-slit screen, the bullet either has hit this screen, or if not, has passed through with a new state $p(\, \cdot \!\mid y)$, where $y$ is the property that it has not hit the screen. This is classically called conditionalizing the state $p$ to $y$. The new state $p(\, \cdot \!\mid y)$ is defined by $p(x\mid y) =  p(x\wedge y)/p(y)$, as the frequency definition of probability can verify. This filtering to a new state is entirely similar to the filtering of a spin 1 system described earlier, and is the classical equivalent of reduction.
 
Now that we have the classical form of reduction as conditionalization, we can follow our prescription by generalizing from a $\sigma$-algebra to a $\sigma$-complex.                                                                                                                            \\

\noindent\textit{Classcal Mechanics: $\sigma$-algebra $B(\Omega)$}\
                                                                                 
Let $p$ be a state on the $\sigma$-algebra $B(\Omega)$ and  $y\in B(\Omega)$ such that $p(y) \ne  0$.  By a \textit{state conditionalized on $y$}  we mean a state $p(\, \cdot\!\mid  y)$ such that for every $x$ in $B(\Omega)$,  
\[
p(x\mid y) = p(x\wedge y)/p(y).
\]

\noindent\textit{General Theory: $\sigma$-complex $Q$}\
                                                                                                                         
Let $p$ be state on a $\sigma$-complex $Q$ and $y\in Q$ such that $p(y)\ne 0$. By a \textit{state conditionalized on $y$} we mean a state $p(\, \cdot\!\mid  y)$  such that for every $\sigma$-algebra $B$  in $Q$ containing $y$ and every $x\in B$,                                                                                                                                                                          \[
p(x\mid y) = p(x\wedge y)/p(y).
\]
 
In the literature, there exist generalizations of  probability measures and conditional probability to non-commutative algebras, and, in particular, to lattices of projections. (See Beltrametti and Cassinelli \cite{2}.)  In general, it is by no means clear that such a state $p(\, \cdot\!\mid  y)$ either exists or is unique, as is obviously the case for classical mechanics. However, for the quantum $\sigma$-complex $Q(\cH)$ this can be proved:
\\

\noindent\textit{Quantum Mechanics: $\sigma$-complex $Q = Q(\cH)$}\
                                                           
If $p$ is a state on $Q(\cH)$ and $y \in Q(\cH) $ such that $p(y) \ne  0$, then there exists a unique state $p(\, \cdot\!\mid  y)$ conditionalized on $y$. If $w$ is the density operator corresponding to $p$, then $ywy/\tr(ywy)$ is the density operator corresponding to the state $p(\, \cdot\!\mid  y)$.                                                                                                                                
\\

To see that the operator $ywy/\tr(ywy)$ corresponds to the state $p(\, \cdot\!\mid  y)$, note that if $x$ lies in the same $\sigma$-algebra as $y$, then $x$ and $y$ commute, so
\[
\tr(ywyx)/\tr(ywy) = \tr(wxy)/\tr(wy) = p(x\wedge y)/p(y) = p(x\mid y).
\]
For uniqueness, it suffices to consider the case when $x \in B(\cH)$ is a one-dimensional projection. Let $p(\cdot  \mid y)$ be a state conditionalized on $y$, and let $v$ be the corresponding density operator. Let $\vp$ be a unit vector in the image of $x$. We can write $\vp = y\vp + y^\bot\vp$. Then 
\begin{align*}p(x \mid y) &= \tr(vx) = \lra{\vp, v\vp}\\
&=\lra{ y\vp, vy\vp} + \lra{y\vp,vy^\bot \vp} + \lra{y^\bot\vp,vy\vp}
+ \lra{y^\bot\vp,vy^\bot\vp}.
\end{align*}
     Now, $\tr(vy^\bot) = p(y^\bot\mid y) = p(y^\bot\wedge y)/p(y) = 0$, so  $vy^\bot\vp = 0$. Hence,   \[
p(x \mid y) = \lra{y\vp,vy\vp} = \|y\vp\|^2 \tr(vP_{y\vp}) = \|y\vp\|^2 p(P_{y\vp})/p(y),
\]
 since $P_{y\vp} \leq  y$. If   $p'(\, \cdot\!\mid   y\}$  is another state conditionalized on $y$, then $$p'(x \mid y) =  \|y\vp\|^2 p(P_{y\vp})/p(y) = p(x \mid y) ,$$  proving uniqueness.      
                                                                                                                 
The change from $w$ to $ywy/\tr(wy)$ in state preparation or measurement is the general formula for the reduction of state given by the  von Neumann-L\"uders Projection Rule. In the orthodox interpretation this rule is an additional principle that is appended to quantum mechanics. Here it appears as the unique answer to conditionalizing a state to a given property.   (See Postulate IIIa of Bohm \cite{4}.)

The natural definition of applying a symmetry $\sig$ to a conditionalized state $p(\, \cdot\!\mid  y)$ is given by \[
p_\sigma(x \mid y) = p(\sig^{-1}(x)\mid \sig^{-1} (y)).
\]  

\ssn{\bf Classical and Quantum Conditional Probability}

 In the well-known paper \cite{9}, Feynman  writes that the basic change from classical to quantum mechanics lies in the revision in the probability rule called the Law of Alternatives, \\ $p(a\mid c) = \sum_i p(a\mid b_i) p(b_i\mid c)$ for disjoint $b_i $, to the quantum law that  $\lra{\alpha\mid \beta} = \sum_i\lra{\alpha\mid \beta_i}\lra{\beta_i\mid \ga}$,  giving an additional interference                                                                                 term.
 
We agree that this is an important difference in the two theories. However, we shall derive it from what we consider the more basic difference, that between intrinsic and extrinsic properties.

 Let $y_1, y_2, \cdots$ lie in a $\sigma$-algebra with $y_i\wedge y_j= 0$ for $i  \ne  j$, and let $y=\bigvee y_i$. Then
\\
                                                          
\noindent\textit{Classical Mechanics:}
\begin{align*}
p(x \mid y) &= p(\mathsmaller{\bigvee} (x\wedge y_i))/p(y)\\
             & = \sum (p(x\wedge y_i)/p(y_i))\cdot  (p(y_i)/p(y))\\
             & = \sum   p(x\mid y_i)p(y_i\mid y),
                \end{align*}                                                                                                     
\emph{The Law of Alternatives} in classical probability theory.

On the other hand,  by Section 7(i), we have
\\

\noindent\textit{Quantum Mechanics:}
\begin{align*}
p(x \mid y)& = \tr(ywyx)/\tr(wy)\\
&=\tr(\mathsmaller{\bigvee}_{i,j} y_i wy_jx)/\tr(wy)\\
&=\sum \tr(y_i w y_ix)/ \tr(wy) + \sum_{i\ne j}\tr(y_i wy_j x)/\tr(wy)\\
&=\sum p(x\mid y_i)p(y_i\mid y) +\sum_{i\ne j}\tr(y_i wy_jx)/\tr(wy).
\end{align*}
This shows that in condionalizing for the extrinsic properties of quantum mechanics an interference term must be added to the classical law of alternatives.   

 \ssn{\bf Conditionalizing on Several Properties}

There is a different kind of preparation of state, one which leads to a mixed state. This occurs when, instead of all but one of the beams being blocked, as in Section7(i), the  beams are allowed to pass through the filter, while being registered. For instance, \cite{8}  describes a version of the two-slit experiment in which the particle scatters high frequency photons that register which slit the particle passed through. In this case, the property $y_1$ of passing through slit 1 is true or the property $y_2$  of passing through slit 2 is true, so that the state of the particle is either the conditional state $p(\, \cdot\!\mid   y_1)$ or the state $p(\, \cdot\!\mid   y_2)$.

If we consider an ensemble of particles, then each of the particles in the ensemble will be in the state $p(\, \cdot\!\mid   y_i)$ with probability $p(y_i)$, for $i =1,2$, so that the ensemble is in the mixed state $p(y_1)p(\, \cdot\!\mid   y_1) + p(y_2)p(\, \cdot\!\mid   y_2)$. Thus, by registering the results of passage through each of the two slits, we restore the classical Law of Alternatives.

For a single particle, the same mixed state describes its predicted state upon passage through the registering two-slit screen. However, upon actual passage through the registered slits, the state is either $p(\, \cdot\!\mid   y_1)$ or $p(\, \cdot\!\mid   y_2)$.  We may say that even after the passage, the state of the particle for an experimenter who is not aware of the registered result the state remains the mixed state. In this regard, the mixture has a similar interpretation as in the classical case, viz., the ignorance interpretation of mixtures.    

 A measurement of an observable is the most familiar example of conditionalizing with respect to several properties. If the observable has a spectral decomposition $\sum a_i P_i $, then measuring the observable amounts to registering the values of the properties given by the $P_i $. The interaction algebra $B$  is the $\sigma$-algebra generated by the $P_i$. 

We now formulate this notion of conditioning with respect to several conditions. 
Given a system with $\sigma$-complex $Q$ and disjoint elements $y_1,y_2 , \dots$  in a  common $\sigma$-algebra in $Q$ with $\bigvee y_i=1$, and a state $p,$ we define the state conditionalized on $y_1,y_2 , \dots$  to be                                                                                                                                                                                                                                                       $p(\, \cdot\!\mid   y_1,y_2 , \dots  ) =  \sum p(y_i)p(\, \cdot\!\mid   y_i)$. We shall also write this more succinctly as $p(\, \cdot\!\mid  B)$, the state conditionalized on the interaction algebra $B$, the $\sigma$-algebra generated by the $y_i$.
  
For quantum mechanics, with $Q = Q(\cH)$,  if $w$ is the density operator corresponding to the \\ state $p$:         \[p(\, \cdot\!\mid  B )  =  \sum \tr(wy_i)(y_iwy_i/\tr(wy_i)) =  \sum y_iwy_i,
\] so that for each $x$ the probability  $p(x\mid B)  =   \sum \tr(y_iwy_ix)$. This gives the state of an ensemble without selection. (See Postulate IIIb of Bohm \cite{4}.)  

The natural definition for applying a symmetry to the conditioned state is given by \[
p_\sig(x \mid B) = p(\sig^{-1}(x)\mid \sig^{-1}B).
\]
Note that the non-uniqueness of the decomposition of a degenerate density operator into pure states causes no problems in this interpretation. This is because mixed states arise as mixtures of given pure states in the conditionalization from an experiment or the evolution of the mixture. The           $\sigma$-algebra $B$  generated by the $y_1,y_2 , \dots$  is simply the current interaction algebra of the $\sigma$-complex, and is always given to us as part of the interaction.             
                                             
The fact that degenerate density operators do not have a unique decomposition into pure states has led some to put mixed and pure states on an equal footing, and to deny them the role as mixtures. This puts the cart before the horse, and ignores the historical development of the concept of mixed states. Mixtures of pure states were in long use in quantum mechanics (as well as in classical statistical mechanics) when von Neumann introduced the invariant formulation of a mixed state as a density operator. The use of the density operator has the advantage of allowing the introduction of the abstract notion of mixed state, without requiring the explicit mention of any basis of pure states, which could be recovered in the non-degenerate case. For us, however, in any interaction (and subsequent evolution) the interaction algebra is always given, which yields a unique decomposition of the mixed state as a mixture of pure states even in the degenerate case.

\section{Reconstructing the $\sig$-Complex $Q(\cH)$}

We saw in Section 1 that if we restrict ourselves to classical experiments, then the $\sigma$-complex of interaction algebras can be imbedded into a $\sigma$-algebra. On the other hand, the 40 quantum triple experiments yield a $\sigma$-complex that cannot be so imbedded. Thus, increasing the set of experiments has changed the structure of the $\sigma$-complexes of systems. It may then be possible that a sufficiently comprehensive family of experiments may force the structure of the $\sigma$-complex $Q$ to be isomorphic to $Q(\cH)$. In this section we shall see that this is indeed the case. 

The result is based on the paper Reck, Zeilinger, Bernstein, Bertani \cite{19}. The interactions arise from a composition of interferometers.  First, Mach-Zender interferometers together with beam splitters allow one to construct $Q(\cH_2)$, where $\cH_2$  is a two-dimensional Hilbert space.
A standard theorem, which allows one to decompose $n$-dimensional unitary operators as a product of two-dimensional ones, is then used to treat the $\sigma$-complex of higher dimensional Hilbert spaces.

We outline the construction in \cite{19} (from which the diagrams below are copied). The experimental realization of a general two-dimensional unitary matrix is obtained by a Mach-Zender interferometer consisting of two mirrors, two 50-50 beam splitters, an $\om$-phase shifter, and  a $\vp$-phase shifter at one output port: 
\[
\begin{picture}(200,120)
\put(10,10){\includegraphics{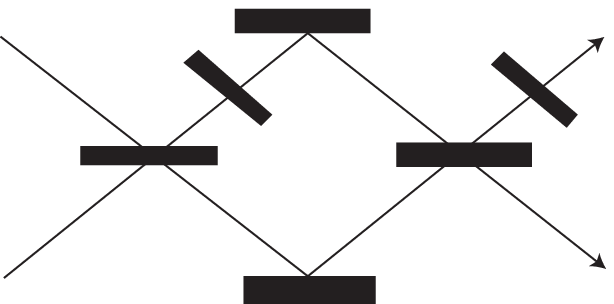}}
\put(0,5){$2$}
\put(188,5){$2'$}
\put(0,90){$1$}
\put(188,90){$1'$}
\put(55,80){$\om$}
\put(140,80){$\vp$}
\end{picture}\]

This device transforms the input state with modes $(k_1,k_2)$ into the output state with modes $(k'_1,k'_2)$, which are related by the unitary matrix: 
\[
\bpm k^\prime_1\\ k^\prime_2 \epm = \bpm e^{i\vp}\sin \om &e^{i\vp} \cos\om\\
\cos\om &-\sin \om\epm \bpm k_1\\ k_2\epm.\]   
We can then realize all 2-dimensional unitary matrices by varying the phase shifters.

To treat $n\times n$ unitary matrices, the authors in \cite{19} show  how to eliminate the off-diagonal element $u_{jk}$ of a unitary matrix $U$ by multiplying $U$ by the matrix $T_{jk}$ which is obtained from the $n\times n$ identity matrix $I$ by replacing the $(jj), (jk), (kj), (kk)$ entries by the entries of a matrix of the above 2-dimensional unitary form. This inductively results in the product
\[
 U T_{nn-1}T_{nn-2} \cdots  T_{32} T_{31} T_{21} = D
\]
 where $D$ is a diagonal unitary matrix with diagonal entries of modulus 1.  Hence 
 \[
U = DT^\dagger_{21}T^\dagger_{31}T^\dagger_{32} \cdots  T^\dagger_{n1}T^\dagger_{n2} \dots T^\dagger_{nn-1}.
\]
                                                                                                                                                                                                                                
We now combine copies of the above interferometers so that the outputs of one are the inputs of the succeeding one, corresponding to the above product of the      $T^\dagger_{jk}$ matrices, followed by n phase shifters to account for the matrix $D$. The result is a device which realizes the matrix $U$. For instance, for $n = 3$, we have:      

\begin{figure}[h]
\[ 
\begin{picture}(400,80)
\put(10,10){\includegraphics[scale=.9]{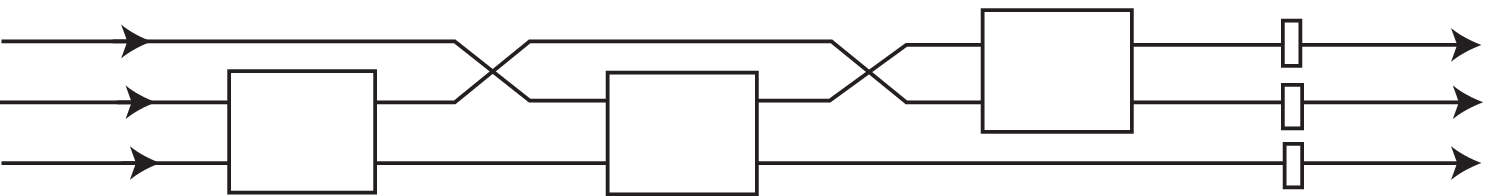}}
\put(80,24){$T^\dagger_{32}$}
\put(180,24){$T^\dagger_{31}$}
\put(280,40){$T^\dagger_{21}$}
\put(360,55){$-\alpha_1$}
\put(360,40){$-\alpha_2$}
\put(360,25){$-\alpha_3$}
\put(350,0){$D$}
\end{picture}
\]
\centerline{(Each box represents an interferometer of the above type.)}\end{figure}

\noindent To realize an $n$-dimensional Hermitean matrix $A$, we use additional beam splitters to superpose those beams that correspond to the same eigenspace of $A$, and then add detectors for the resulting beams. The use of beam splitters to superpose beams is well-known. (See e.g. Zukoowski, Zeilinger, and Horne [23].)        
                                                                                                                       
This is a pr\'ecis of the construction in \cite{19}. It allows us to realize every element of $Q(\cH)$, where $\cH$ is an $n$-dimensional complex Hilbert space. What is significant is that we can also realize the $\sigma$-complex structure of $Q(\cH)$. To see this it suffices to consider the two Boolean operations of complementation $x^\bot$ and join $x \vee y$. The output for a projection $x$ consists of two beams, labeled the 1-beam and the 0-beam according to the eigenvalues of  $x$. The operation of complementation $x^\bot$ requires only a transposition of the 1 and 0 labels. The join $x \vee y$ of two projections corresponds to superposing the two 1-beams of $x$ and $y$ . These two operations suffice to define all the Boolean operations, and therefore the $\sigma$-complex structure of $Q(\cH)$. Note that this realization of the $\sigma$-complex of properties via the different $\sigma$-algebras generated by the outcomes of interferometer experiments follows the general prescription given in Section 1 for defining the $\sigma$-complex of properties of a system by means of the different $\sigma$-algebras of events defined by the experimental outcomes.

It is instructive to contrast the simple experimental counterparts to the $\sigma$-complex structure with the lattice structure of the set of projections. We know of no  corresponding experimental realization to the lattice join (or meet) of two non-commuting projections. This is due to the difficulty of relating the eigenspaces of two non-commuting operators to the eigenspaces of their sum (or, for projections, to their union), while for commuting operators there is a simple relation. It is this difficulty that is alluded to in our earlier quotations from Varadarajan \cite{21} and Birkhoff and von Neumann \cite{3} in the introduction.  

We have seen that if we can in principle form arbitrarily large networks of interferometers, then we can realize the $\sigma$-complex $Q(\cH)$ for Hilbert spaces of all finite dimensions. The single minimal space $\cH$ for which $Q(\cH)$ realizes \textit{all} the interferometer experiments, and hence contains all finite dimensional Hilbert spaces, is an infinite dimensional separable pre-Hilbert space, i.e. an inner product space $\cH,$ whose completion forms a separable Hilbert space $\cH_\omega.$ To see this note that $\cH$ may be construed as the space of all complex sequences $\{a_i\}$ that are non-zero for only finite many $i,$ with inner product $\langle \{a_i\},\{b_i\}\rangle = \sum a_i b_i.$

Thus in the infinite dimensional case we must add ideal elements which are limits of sequences of realized elements. We cannot expect to realize $Q(\cH_\omega)$ via experiments without adding limits since the world itself may be finite. This is similar to the use of probability in physica as an ideal limit of relative frequency for longer and longer sequences of experiments. Of course, even the above realization of $Q(\cH)$ in the finite dimensional case is an idealization, since it requires $\omega$-phase shifters for arbitrary real $\omega$, in $[0,2\pi]$.

We may now extend the result $Q(\cH_1)\oplus Q(\cH_2) \simeq Q(\cH_1 \otimes \cH_2)$ of Section 4 to the infinite dimensional case. 

The fact that $\cH$ is the \emph{minimal} space such that $Q(\cH)$ is realized by the above interferometry experiments highlights the open-ended nature of our reconstruction. If we restrict ourselves to experiments of classical physics, then the $\sigma$-complex reduces to a $\sigma$-algebra, and the concepts lead to classical physics. If we add the forty triple experiments, the resulting $\sigma$-complex cannot be imbedded into a $\sigma$-algebra. If we allow for the interferometry experiments of this section, then $Q$ must take the form $Q(\cH).$ It thus suffices to consider these interferometry experiments to realize the structure of quantum physics. We may then apply the resulting theory to general interactions\footnote{Historically, of course, it was not such interferometry experiments, but rather spectroscopic experiments that lead Schr\"{o}dinger to his equation.}. As we have emphasized throughout the paper, the special nature of experiments, with the macroscopic apparatus, plays no role in the theory. Any appropriate decoherent interaction gives rise to isomorphic $\sigma$-algebras for the two systems. Experiments do play the pragmatic role of allowing us to become cognizant of a sufficient number of interactions to help deteremine the theory. 

It is possible that other experiments may require a different realization of the $\sigma$-complexes. For instance, if we consider systems which satisfy \emph{superselection rules} (see e.g. [2]), then the $\sigma$-complex $Q$ has a non-trivial $\sigma$-algebra which is common to all the $\sigma$-algebras $B$ in $Q$. In this case $Q$ is not of the form $Q(\cH),$ but is a sub-$\sigma$-complex of $Q(\cH)$. $\cH$ takes the form of a direct sum $\oplus \cH_i$ of Hilbert spaces with the pure states forced to lie in a factor $\cH_i$.

\section{From Quantum Physics to Classical Physics}
    
With the description in Section 4 of the $\sigma$-complex of combined systems, it is possible to treat the statistics of a large number of particles such as macroscopic bodies. This is, of course, a major subject in quantum statistics, and we shall not venture there. However, we wish to say a few words on how the $\sigma$-complex of quantum mechanics tends to a classical $\sigma$-algebra with an increasing number of particles, so that the quantum system becomes effectively classical.    
                  
We shall adapt a remark in Finkelstein \cite{10} for this purpose. Let $S$ be an ensemble of $n$ non-interacting copies of a system $S_i$, $i = 1,2,\dots ,n$, with  $\sig$-complex $Q(\cH_i)$. Then $S$ has the $\sigma$-complex \[
Q(\cH_i)\oplus Q(\cH_2) \oplus\cdots \oplus Q(\cH_n) \simeq Q(\cH_1\otimes \cH_2\otimes\dots \otimes\cH_n).
\]
 Suppose each $S_i$  is in the pure state $\vp$. Then $S$ is in the state $\Phi = \vp\otimes \vp\otimes\dots\otimes\vp$. Consider the observable $\bA$ of $S$ which is the average of the same observable $A$ of each $S_i $:
  \[
\bA = (A\otimes I\otimes \dots\otimes I + I\otimes A\otimes \dots \otimes I + \dots  + I\otimes I\otimes\dots \otimes A)/n. 
\]
We recall that the the uncertainty $\Delta R$ of an operator $R$ is the square root of the variance: \newline $(\Delta R)^2  = \Exp((R - \Exp R)^2)$.  Hence,  
\begin{align*}
 (\Delta \bA)^2&= \lra{\Phi, (\bA-\Exp \bA)\Phi}\\
                                      &= (1-1/n)\lra{\vp,(A-\Exp A )\vp}^2 +(1/n)\lra{\vp,(A-\Exp A )^2\vp}\\
                                      &= (\Delta A)^2/n.
\end{align*} 
Hence, if  $$\bB  =(B\otimes I\otimes \dots \otimes I + I\otimes B\otimes \cdots I + \dots  + I\otimes I\otimes \dots \otimes B)/n$$ is another such averaged observable, then for the commutator $[A,B]$ we have               \[
\Delta[\bA,\bB] = \Delta[A,B]/n. 
\] 
Thus, $\lim_{n\to\infty} \Delta[\bA,\bB]  = 0$.   It follows that the averaged observables of  $S$ all commute in the limit, and so the $\sigma$-complex of $S$ becomes essentially a $\sigma$-algebra for very large $n$, as in a macroscopic body.  

This calculation was made under the assumption that $S$ is an ensemble of non-interacting replicas of one particle. In a real body the states and observables need not be identical. Without going into details, it is possible to give conditions on the allowed variation of the the states of the particles and the averaged observables so that  $\Delta[\bA,\bB]$  still tends to zero with increasing $n$. In any case, the result is at least suggestive that in a real body, the $\sigma$-complex of S will be very close to a $\sigma$-algebra.                  
                         
The change in dynamics accompanying the move from the Hilbert space $\cH$ to the phase space $\Omega$ has been well-studied. In essence, the quantum bracket $\frac{i}{\hbar}[X,Y]$ is replaced by the Poisson bracket $\{X,Y\}$, so that the von Neumann-Liouville equation  $\partial_tw_t = -\frac{i}{\hbar} [ H, w_t ]$ is replaced by the classical Liouville equation $\partial_t f_t = -\{ H, f_t\}$. (See Faddeev and Yakubovskii [7], for example.) We saw in Sections 5 and 6 that the lack of sufficient structure of a $\sigma$-algebra did not allow us to derive the classical dynamics from the automorphisms of $B(\Omega)$, whereas we could do so in the quantum case $Q(\cH)$. We can see now how it is possible to recover the classical dynamical equation by an excursion into the quantum structure $Q(\cH)$. 

 \section{Interpreting and Resolving  Quantum Paradoxes.}

 \ssn{\bf The K-S Paradox and the Projection Rule}

We have already applied this reconstruction to treat several issues in the interpretation of the formalism. One of these, the Kochen-Specker Paradox, which showed that the assumption that all properties are intrinsic leads to a contradiction, was the motivation for introducing the $\sigma$-complex of extrinsic properties. Conversely, assuming the relational nature of properties resolves this paradox. Another issue, discussed in Section 7, is the nature of reduction and the von Neumann-L\"uders Projection Rule, which here appears as the counterpart to classical conditionalizing, not as an ad hoc addition to quantum theory.  We now consider a number of other controversial questions from the literature. 

\ssn{\bf Wave-Particle Duality}
 
We discuss wave-particle duality in the context of the two-slit experiment. Let $y_1$ and $y_2$ be the projections of position in the regions of the two slits $\delta_1$ and $\delta_2.$ Then $y_1\vee y_2$ is the projection of position for the union $\delta_1 \cup \delta_2.$ Let $x$ be the property of position in a local region $\Delta$ on the detection screen. 

If passage through each of the two slits is registered, then the Law of Alternatives of Section 7(ii) tells us that $p(x|y_1\vee y_2)=p(x|y_1)p(y_1|y_1\vee y_2) + p(x|y_2)p(y_2|y_1\vee y_2),$which, in the case of symmetrical positioned slits, is propotional to the sum $p(x|y_1)+p(x|y_2)$ of the probabilities of passage through the individual slits, just as in the classical case. 

In the case where the passage through the two slits by the quantum particle is not registered, we have shown in Section 7(ii) that there is an additonal interference term \[[tr(y_1wy_2)x)+tr(y_2wy_1x)]/ tr(w(y_1\vee y_2)).\] Note that if $x$ and $y_1$ and $y_2$ commute, this interference term vanishes. This happens if the detector is right next to the two-slit screen. If the detector is a distance from the two-slit screen, then the particle undergoes free flight evolution $\sigma_t,$ so $\sigma_t(y_i)=u_ty_iu^{-1}_t$ no longer commutes with $x$, giving rise to the non-zero interference term.
 
An explanation of the interference effect that is often given is that the particle is, or acts as, a pair of waves emanating from the slits, which exhibit constructive and destructive interference effects. This was, of course, the explanation for Young's original experiment with the classical electromagnetic field. For individual quantum particles however, it leads to the paradoxical effect that the wave suddenly collapses to a local region  at the detection screen.  

The explanation given here is a different one. A system forms a localized particle if there is a position operator for the system, so that a measurement of position detects the system at a localized region in space. Until the position is measured                                                                                                                                                            the position has no value, since position in a region is an extrinsic property. We may view the two-slit screen as a preparation of state for the particle, for which the position is conditionalized, or reduced, to the region $\delta_1\cup\delta_2 $. This reduction is not a position measurement, since $\delta_1\cup\delta_2$ is not a localized region (as it would be for a single-slit screen). It is only at the detection screen, where the particle, in interaction with the screen, is reduced to the local region $\Delta$, that its position has a value.

The question of why the particle shows the interference effects of a wave is answered in Section 6, where the evolution of the quantum particle was defined by a trajectory in the space $\Aut(Q)$. This yielded the Schr\"{o}dinger equation, which is a wave equation. On the other hand, a trajectory in the phase space of a classical particle passing through a two-slit screen is governed by the classical Liouville equation, without any wave properties. Thereby, the wave-like properties of a quantum particle are explained by the extrinsic character of of its properties.  

\ssn{\bf The Measurement Problem}                                                                                                                         

The Measurement Problem refers to an inconsistency in the orthodox interpretation of quantum measurement. The interpretation assumes that an isolated system undergoes unitary evolution via Schr\"{o}dinger's equation. We quote from Bohm \cite[Chapter XII]{4}:                                                                                                                        
\begin{quote}``If time evolution is a symmetry transformation, then the mathematical structure (in particular the algebraic relations) of the algebra of observables does not change in time; this means that the physical structure is indistinguishable at two different points in time. Our experience shows that there are physical systems        that have this property and in fact it is this property that defines the isolated systems. Thus isolated physical systems do not age, an absolute value of time has no meaning for these systems, and only time differences are accessible to measurement. Irreversible processes do not take place in isolated physical systems defined as above.''
\end{quote}
Accordingly, in the orthodox interpretation, for  a measurement of an observable $A$ of a system $S$ by an apparatus $T$, the total system $S+T$, which is assumed to be isolated, undergoes unitary evolution. 

 We outline the standard description of an ideal measurement. Suppose the spectral decomposition of an observable is $A =\sum a_i P_i$, where each $P_i$ is a one-dimensional projection with eigenstate $\vp_i$. The apparatus is assumed to be sensitive to the different eigenstates of $A$. Hence, if the initial state of $S$ is $\vp_k$ and the apparatus $T$ is in a neutral state $\psi_0$, so that the state of $S+T$ is $\vp_k\otimes\psi_0$, then the system evolves into the state $\vp_k\otimes \psi_k$, where the $\psi_i$ are the states of the apparatus co-ordinate corresponding to the states  $\vp_i$ of the system. By linearity, if $S$ is in the initial state $\vp=\sum a_i\vp_i$, then $S+T$ evolves into the state $\Gamma=\sum a_i \vp_i\otimes \psi_i$. The intractable problem for the orthodox interpretation is that the completed measurement gives a particular apparatus state $\psi_k$, indicating that the state of $S$ is $\vp_k$, so that the state of the total system is $\vp_k\otimes \psi_k$, in contradiction to the evolved state  $\sum a_i \vp_i\otimes \psi_i$.  We may also see the reduction from the viewpoint of the conditionalization of the states. If the state $p$ of $S+T$ just prior to measurement is $P_\Gamma,$ then after the measurement it is the conditionalized state \[p(\ \cdot \ |P_{\phi_k}\otimes I \wedge I\otimes P_{\psi_k})=(P_{\phi_k}\otimes I \wedge I \otimes P_{\psi_k}) P_\Gamma (P_{\phi_k} \otimes I \wedge I \otimes P_{\psi_k})/ tr((P_{\phi_k} \otimes I \wedge I \otimes P_{\psi_k}) P_\Gamma )) = P_{\phi_k \otimes \psi_k}.\] Hence, the new conditionalized state of $S+T$ is the reduced state $\phi_k \otimes \psi_k.$
 
The orthodox interpretation then has to reconcile the unitary evolution of  $S+T$ with the measured reduced states of $S$ and $T$. The present interpretation stands the orthodox interpretation on its head. We do not begin with the unitary development of an isolated system, but rather with the results of a measurement, or, more generally, of a decoherent interaction. In fact, the original motivation for forming a $\sigma$-complex of properties was via the set of measured, and hence reduced, properties which form the current interaction algebra. For us, it is the conditions under which dynamical evolution occurs that is to be investigated, rather than the reduced state.
We cannot take for granted what is assumed in the orthodox interpretation, as in the above quotation, that an isolated system evolves unitarily.  So we must answer the question whether in a measurement the $\sigma$-complex structure of $S+T$ undergoes a symmetry transformation at different times of the process. As Section 6 showed, this is formalized as the condition for the existence of a representation $\sigma:\R\to \Aut(Q)$.
                                                   
It is easy to see, however, that in the process of a completed measurement or a state preparation there are two distinct elements of $Q(\cH) (=Q(\cH_1 \otimes \cH_2))$ at initial time 0 which end up being mapped to the same element at a later time $t$. We have seen that an initial state $\vp \otimes \psi_0$ results in a state $\vp_k\otimes \psi_k$, for some $k$. However, $\vp_k\otimes \psi_0$ also results in the state $\vp_k \otimes \psi_k$. If we choose the state $\vp$ to be distinct from $\vp_k$, then the two elements $P_{\vp\otimes \psi_0}$ and $P_{\vp_ k\otimes \psi_0}$  of $Q(\cH)$ both map to the same element $P_{\vp_k \otimes \psi_k}$.  However, any automorphism $\sig_t$ is certainly a one-to-one map on $Q$, so the measurement process cannot be described by a representation  $\sig:\R\to \Aut(Q)$, and hence a unitary evolution.                      
                                                                                                    
In our interpretation, the Measurement Problem is thus resolved in favor of reduction rather than unitary evolution. The point can be made intuitively that points of absolute time do exist in a measurement and also in state preparation, namely the point (or, better, small interval) of time at which reduction takes place. If for instance, we consider a Stern-Gerlach experiment with a state preparation in which a filter registers the passage of a particle through one of several slits, before the particle reaches a detection screen, then the interval of time of passage through the slit, in which the state of the particle is reduced, is such an absolute point of time: the state after passing through the slit is the conditionalized state, whereas before it is not.

Time and its passage is a problematic concept in physics, so to reinforce the point we shall give another example, in which time homogeneity is tied to spatial symmetry. Consider a particle resulting, say, from decay in which its state has spherical symmetry. Assume that the particle is initially at the center of a spherical detector system. During the passage of the particle until it hits the detector, the combined system of particle and detector is spherically symmetric and time homogeneous. At the moment of registering the impact on a local region ofthe detector, the system loses both its isotropy in space and its time symmetry. If it is difficult to argue against this breaking of space symmetry in favor of a particular direction, it seems to us to be equally hard to gainsay the breaking of time symmetry at the moment this non-isotropy occurs.                      
                                                                                                     
For a composite system it is not only outside forces that can break symmetry, but internal interactions. As opposed to the quotation of Bohm\cite{4} above, we believe that symmetry-breaking processes do take place in isolated compound systems with internal decoherent interactions during reduction of state. To argue that nevertheless symmetry has not been broken for the combined system is to favor the theoretical formalism ahead of the facts on the ground. It is notable that with this interpretation the system consisting of the universe as a whole, for which there are no external systems, acquires reduced or, as we say, conditionalized states as a result of the interactions of component systems. 

Note that our alternative term \textit{interactive property} is more appropriate here than \textit{extrinsic property.} The reduction of the state to $\phi_k \otimes \psi_k$ happens for the composite system $S_1+S_2$ because of the interaction of the component systems $S_1$ and $S_2$ which are internal to $S_1+S_2$ rather than an interaction of $S_1+S_2$ with an external system.

\ssn{\bf The Einstein-Podolsky-Rosen Experiment} 

 We shall discuss the EPR phenomenon in the Bohm form of two spin $\frac{1}{2}$ particles in the combined singlet state $\Ga$ of total spin 0. Suppose that in that state the two particles are separated and the spin component $s_z$ of particle 1 is measured in some direction $z$. That means that the observable  $s_z \otimes I$ of the combined system is being measured. 
 
 Let $P_z^\pm   = \frac{1}{2} I \pm  s_z$ . We have the spectral decomposition
       \[
s_z \otimes I = \mathsmaller{\frac{1}{2}} P_z^+ \otimes I + (-\mathsmaller{\frac{1}{2}})P_z^-\otimes I ,   
\]
so the interaction algebra $B  = \{0,1, P_z^+ \otimes I, P_z^-\otimes I\}.$  We expand the singlet state   
\[
\Ga= \sqrt{\mathsmaller{\frac{1}{2}}} (\phi^+_z \otimes \psi^-_z - \phi^-_z \otimes \psi^+_z)
,\] where   $P^\pm_z \phi^\pm_z = \phi^\pm_z$ and $P_z^\pm \psi_z^\pm = \psi_z^\pm$. Thus, if particle 1 has spin up, the state $p(\, \cdot\!\mid   P_z^+\otimes I)$ of the system is, by Section 7(i), given by 
\[
p(\, \cdot\!\mid   P^+_z\otimes I)  =(P^+_z \otimes I) P_\Ga (P^+_z \otimes I)/\tr((P^+_z \otimes I)P_\Ga) = P_{\phi^+_z \otimes \psi_z^-}.  
\]
This is, of course, equivalent to projecting the vector $\Ga$ into the image of  $P^+_z\otimes I$:                                                                                                                          \[
P^+_z \otimes I (\Ga)  = \sqrt{\mathsmaller{\frac{1}{2}}} (\phi^+_z \otimes \psi^-_z).
\]  
  Similarly, if particle 1 has spin down the state $p(\, \cdot\!\mid   P_z^-\otimes I)$ is given by the vector                                                                                                   \[
P^-_z\otimes I(\Ga)=\sqrt{\mathsmaller{\frac{1}{2}}}(\phi^-_z \otimes \psi^+_z).
\]
This shows that if $s_z$ is measured for particle 2, it is certain to have opposite value of $s_z$ for particle 1. It does \textit{not} mean that after $s_z$ is measured for particle 1, then $s_z$ has a value for particle 2. The properties $I \otimes P^+_z$ and $I\otimes P^-_z$ do not lie in the interaction algebra $B=\{P^+_z\otimes I, P^-_z\otimes I, 0, 1\}$, and so have no value. The spin components are extrinsic properties of each particle, which do not have values until the appropriate interaction. To claim otherwise is to revert to the classical notion of intrinsic properties. 

This is a necessary consequence of our interpretation, but it also follows from a careful application of standard quantum mechanical principles. For after the measurement of $s_z$ on particle 1 gives a value of $\frac{1}{2}$, the state of the combined system is $\phi^+_z\otimes\psi^-_z,$ which is an eigenstate of $I\otimes s_z.$ Born's Rule implies that an eigenstate of an observable will yield the corresponding eigenvalue as value only if and when that observable is measured. 

The situation is entirely similar to the unproblematic triple experiment. A triple experiment on the frame $(x,y,z)$ yields the interaction algebra $B_{xyz}$. If $S^2_z=0$, then $S^2_x=S^2_y=1.$ If $(x^\prime, y^\prime, z)$ is another frame, then it is also the case that $p(S^2_{x^\prime}=1|S^2_z=0)=1,$ so that $S^2_{x^\prime}$ is certain to have the value 1 if the triple experiment on the frame $(x^\prime, y^\prime, z)$ is performed. But $S^2_{x^\prime}$ does not have a value unless and until that experiment is carried out since $S^2_{x^\prime}=1$ does not lie in the interaction algebra $B_{xyz}$.
                                                      
We have not in this discussion mentioned a word about special relativity. Indeed, the spin EPR phenomenon has nothing to do with position or motion and is independent of relativistic questions. However, EPR with space-like separated particles has been used to put in question the full Lorentz invariance of quantum mechanics. This is replaced by a weaker notion that EPR correlations cannot be used for faster than light signaling. We believe that Lorentz invariance is a fundamental symmetry principle, which gives rise to basic observables, and is not simply an artifact of signaling messages between agents.

The relativistically invariant description of the EPR experiment is that if experimenters $A_1$ and $A_2$ measure particles 1 and 2, and the directions                                                                                                                             of spin in which they are measured are the same, then an experimenter $B$  in the common part of the future light cones of $A_1$ and $A_2$ will find that the spins are in opposite directions. 
                                                                                                                              
Now that we have studied what EPR actually says, we shall treat the question of how correlations can exist between the different directions of spins of two particles when such spins cannot simultaneously have values.

To set the stage for EPR, we again first consider the triple experiment. For a spin $1$ particle the proposition $S^2_z=1$ defines the same projection in $Q(\cH)$ as the proposition 
\begin{equation}
	S^2_x=0 \leftrightarrow S^2_y=1
\end{equation}
If we perform the $(x,y,z)$ triple experiment with interaction algebra $B_{xyz}$ and find that $S^2_z=1,$ then we can check that either $S^2_x=0$ and $S^2_y=1$ or $S^2_x=1$ and $S^2_y=0,$ so that (1) is true. However, for the orthogonal triple $(x^\prime,y^\prime,z)$
\begin{equation}
	S^2_{x^\prime}=0\leftrightarrow S^2_{y^\prime}=1
\end{equation}
is the same projection as (1) and so is also true. But $S^2_{x^\prime}$ and $S^2_{y^\prime}$ do not lie in the interaction algebra $B_{xyz},$ and so have no truth value unless and until the $(x^\prime,y^\prime,z)$ triple experiment is performed. Thus, the correlation $(2)$ is true without its component properties $S^2_{x^\prime}$ and $S^2_{y^\prime}$ having truth values.

Now consider the EPR experiment. We have seen in Section 4 tha $S=0$ is the same projection as $(S_z=0) \wedge (S_x=0),$ and $S_z=0$ and $S_x=0$ are in turn respectively the same projections as 
\begin{equation}
	s_z\otimes I = \frac{1}{2} \leftrightarrow I \otimes s_z = -\frac{1}{2}
\end{equation} and
\begin{equation}
	s_x \otimes I = \frac{1}{2} \leftrightarrow I \otimes s_x = -\frac{1}{2}.
\end{equation}
If the projections $S_z=0$ and $S_x=0$ are true, then so are the correlations (3)and (4) since they define the same projections. As in the triple experiments, we see that these correlations subsist simultaneously, even though the spins $s_z$ and $s_x$ for each particle cannot have values simultanously. Thus, the existence of seemingly paradoxical EPR correlations in different directions can be understood via the logic of extrinisic properties. 
                                                                                                            
In summary, the extrinsic properties of a $\sigma$-complex may have relations subsisting among its elements because of general laws of physics, such as conservation laws, which are timeless and independent of particular interactions. The $\sigma$-complex structure accommodates such relations in the form of compound formulas such as $(3)$ and $(4)$, which are true, even when the constituent parts do not have truth values.  This fact allows us to interpret the EPR phenomenon in a fully relativistically invariant way. For extrinsic properties a compound property may have truth values even when the component parts do not.  \\

\section{On the Logic of Quantum Mechanics}

As we have stressed thoughout this paper, the major transformation from classical to quantum physics in this approach lies not in modifying the basic classical concepts such as state, observable, symmetry, dynamics, combining systems, or the notion of probability, but rather in the shift from intrinsic to extrinsic properties.

Now properties, whether considered as predicates or propositions, are the domain of logic. Boolean algebras correspond to propositional logic and $\sigma$-algebras to predicate logic. Hence the change to a $\sigma$-complex of exrinsic properties should entail a new logic of properties. At first sight however, it would appear that the logic of extrinsic properties as elements of a $\sigma$-complex $Q$ is no different than classical propositional logic, since these elements can only be compounded when they lie in the same $\sigma$-algebra in $Q$. This is far from the case; in fact, the difference in logic plays an important role in resolving some of the quantum paradoxes. The underlying reason is that a compound property such as $x \vee y$ may be lie in an interaction algebra and so have a truth value, even though neither $x$ nor $y$ lie in the algebra, and have no truth value.

The logic of extrinsic properties has been sysematically studied in Kochen, Specker \cite{14} and \cite{15}, where a complete axiomatization of the propositional calculus of extrinsic properties is given. Here we shall confine ourselves to pointing out some uses of this logic that appeared in this paper. 

1. The simplest such case is $x\vee x^\perp,$ which equals 1 in $Q$, and so is always true, even though $x$ may have no truth value.\footnote{This is reminiscent of Aristotle's famous sea battle in \textit{De Interpretatione}: ``A sea battle must either take place tomorrow or not, but it is not necessary that it should take place tomorrow neither is it necessary that it should not take place, yet it is necessary that it either should or should not take place tomorrow.''} Thus, for a spin $\frac{1}{2}$ particle, $s_z=\frac{1}{2} \vee s_z=-\frac{1}{2}$ is true simultaneously for all directions $z$, though $s_z$ may have no value.

2. In the two-slit experiment (Section 10(ii)), we saw that it is this lack of truth value that leads to the interference pattern at the detector screen. The source of the interference pattern is not some non-classical probability, but rather the applications of classical Kolmogorov axioms of probability to the logic of extrinsic propeties. The conditional probability $p(x|y)$ is the probability of $x$ given that $y$ has happened and so has a truth value. Therefore the probability $p(x|y_1\vee y_2)$ implies that $y_1\vee y_2$ is true. However, neither $y_1$ nor $y_2$ has happened. We should not expect the classical Law of Alternatives connecting $p(x|y_1\vee y_2)$ to $p(x|y_1)$ and $p(x|y_2)$ to be valid unless $y_1$ and $y_2$ are events that have happened. In that case the Law of Alternatives is in fact valid in quantum mechanics.

3. In the EPR experiment, the singleton state $S=0$ implies that $s_z\otimes I = \frac{1}{2} \leftrightarrow I \otimes s_z = -\frac{1}{2}$ is true for any direction $z.$ In fact, as shown in Section 4 the element $S=0$ equals \[(s_z \otimes I = \frac{1}{2} \leftrightarrow I \otimes s_z = - \frac{1}{2}) \wedge (s_x \otimes I = \frac{1}{2} \leftrightarrow I \otimes s_x = -\frac{1}{2}).\] Thus, the correlation exists in both the $z$ and $x$ directions even though the spins cannot simultaneously have values in these directions. Section 4 shows how general superpositions of states of combined systems may be reformulated as compound statements of this quantum logic.

4. The K-S Paradox in Section 1 can be stated as a proposition that is classically true but false in quantum mechanics. To see this, let $+$ denote exclusive disjuntion. Then $x+y+z+x\wedge y\wedge z$ is true if and only if exactly one of $x, y,$ and $z$ is true.

The statement $\bigvee_{i\leq 40} (x_i+y_i+z_i+x_i\wedge y_i \wedge z_i)^\perp$, where $(x_i,y_i,z_i)$ range over the orthogonal triples of the 40 triple experiments of Section 1 is classically true, but false under a substitutions $x_i \mapsto S^2_{x_i}, y_i \mapsto S^2_{y_i}, z_i \mapsto S^2_{z_i}.$

For two spin $\frac{1}{2}$ particles there is a K-S Paradox which yields a much simpler such proposition in four dimensional Hilbert space: \[ [(x\leftrightarrow y)\leftrightarrow (z\leftrightarrow w)] \leftrightarrow [ (x\leftrightarrow z)\leftrightarrow (y\leftrightarrow w)].\] This classically true proposition is false under the substitution  \[ x \mapsto s_z \otimes I = \frac{1}{2}, \ y\mapsto I \otimes s_z = \frac{1}{2}, \ w \mapsto s_x \otimes I = \frac{1}{2}, \ z \mapsto I \otimes s_x = \frac{1}{2}.\footnote{For details, see J. Conway and S. Kochen, \textit{The Geometry of the Quantum Paradoxes, Quantum [Un]speakables}, R.A. Bertlemann, A. Zeilinger (ed.), Springer-Verlay, Berlin, 2002, 257.}\] Kochen, Specker \cite{13} Theorem 4 shows that every K-S Paradox corresponds to a classically true proposition which is false under a substitution of quantum properties.

\vfill
\noindent	Mathematics Department, Princeton University.\\
Dedicated to the memory of Ernst Specker.\\
This work was partially supported by an award from the John Templeton Foundation.                                                  

\section{Appendix: Summary Table of Concepts}

\begin{center}
{
  \renewcommand\arraystretch{1.1}\begin{tabular}{|l|l|l|l|}
\hline
&General Mechanics&
     Classical Mechanics&
  Quantum Mechanics
 \\  \hline
Properties&  
$\sigma$-complex
 &$\sigma$-algebra
&$\sigma$-complex
  \\
 &$Q = \cup B$, with $B$  a  $\sigma$-algebra& 
\multicolumn{1}{|c|}{$B(\Omega)$} & \multicolumn{1}{|c|}{$Q(\cH\}$} \\ \hline
States &
$p: Q\to [0,1]$ & 
 $p: B(\Omega)\to [0,1]$ &  
$w: \cH\to \cH$  
\\ &$p\mid B$, a probability measure &a probability measure & Density operator\\
&&&$p(x)=\tr(wx)$
\\
 \hline
Pure States   &                
Extreme point  
& %
&%
1 dim operator\\
&of convex set& \multicolumn{1}{|c|}{$\om\in\Om$} &i.e.\ unit  $\phi\in\cH$\\
&&&$p(x)=\lra{x,x\vp}$\\ \hline
Observables&
$u: B(\R)\to Q$ 
& %
   $f:\Om\to \R$
& 
$A: \cH \to \cH$
\\
&homomorphism&Borel function&Hermitean operator\\ \hline 
Symmetries
&$\sigma:Q \to Q$& %
 $h:\Omega\to \Om$&
$u:  \cH\to \cH$\\
&automorphism&
canonical&
unitary or \\
&&transformation &
anti-unitary operator\\
&&&
$\sig(x)=uxu^{-1}$\\\hline
Dynamics&
$\sig:\R\to \Aut(Q)$&
Liouville equation&
von Neumann\\
&representation&
$\partial_t \rho=-[H, \rho]$&
-Liouville equation    \\
&&&                 $\partial_t w_t=-\frac{i}{\hbar} [ H, w_t]  $   \\              \hline           
Conditionalized
&
$ p(x) \to p(x \mid y)$&
$p(x) \to p(x \mid y)$&
 $w\to  ywy / \tr(wy) $\\
 States&
 for $x,y\in B$ in $Q$&
 $=p(x \wedge y)/p(y)$& 
von Neumann\\
&$p(x \mid y)=p(x \mid y)/p(y)$&&
-L\"uders Rule        \\
\hline 
Combined
&
$Q_1 \oplus Q_2$ & $\Omega_1 \times \Omega_2$ & $\cH_1 \otimes \cH_2$\\

Systems & direct sum of  & direct product of & tensor product of \\ 
& $\sigma$-complexes & phase spaces & Hilbert spaces \\ \hline    
\end{tabular}} \end{center}

\end{document}